\documentclass[fleqn,usenatbib]{mnras}

\usepackage{newtxtext,newtxmath}
\usepackage[T1]{fontenc}

\DeclareRobustCommand{\VAN}[3]{#2}
\let\VANthebibliography\thebibliography
\def\thebibliography{\DeclareRobustCommand{\VAN}[3]{##3}\VANthebibliography}

\usepackage{graphicx}	% Including figure files
\usepackage{amsmath}	% Advanced maths commands
\usepackage{tabularx}
\usepackage{caption}
\usepackage{subcaption}
\usepackage{xcolor}

\def\redpen#1{\textcolor{black}{#1}}
\def\blackpen#1{\textcolor{black}{#1}}

\newcommand{\Msun}{M$_\odot$}
\newcommand{\Rsun}{R$_\odot$}
\newcommand{\GAIA}{\textit{Gaia}}
\def\SPSB#1#2{\,\rlap{\textsuperscript{#1}}\textsubscript{#2}}

\title[Mass determination of the eclipsing heartbeat star HD 181793]{Dynamical mass determination and partial eclipses of the heartbeat star HD 181793}

\author[Uronen et al.]{
Laura E. Uronen,$^{1}$\thanks{laura.uronen@gmail.com}
Andrew Collier Cameron,$^{1}$ Thomas G. Wilson$^{1,2}$
\\
$^{1}$SUPA School of Physics \& Astronomy, University of St~Andrews, North Haugh, St~Andrews KY16~9SS, Scotland, UK\\
$^{2}$Department of Physics, University of Warwick, Coventry, West Midlands, CV4 7AL, England, UK
}

\date{Accepted 2024 June 16. Received 2024 May 30; in original form 2023 August 22}

\pubyear{2023}

\begin{document}
\label{firstpage}
\pagerange{\pageref{firstpage}--\pageref{lastpage}}
\maketitle

\begin{abstract}
%An eclipsing, chemically peculiar heartbeat binary is a rare object to begin with. HD 181793, one such system with a bright Am-type primary, is only the second of its kind. 
\textcolor{black}{We identify the bright Am-type star HD 181793 to be a previously-unknown eclipsing, chemically peculiar heartbeat binary, the second of its kind known.}
The system carries an orbital period of $P = 11.47578275 \pm 0.00000055$ days.
%I have used 
We use {\it TESS} photometry and LCOGT NRES radial velocity data to build a \textcolor{black}{self-consistent} orbital model and \textcolor{black}{determine} the fundamental stellar characteristics of the primary. \textcolor{black}{We use} a spectral separation method to unveil the secondary \textcolor{black}{and measure the masses of both stars}. The radial velocity amplitude of the primary, $K_1 = 47.41\SPSB{+0.13}{-0.12}$ km s\textsuperscript{-1}, gives a mass \blackpen{$M_1 = 1.57 \pm 0.01 $ \Msun{}}. The secondary radial velocity amplitude $K_2 = 84.95\SPSB{+0.12}{-0.09}$ km s\textsuperscript{-1} yields a mass ratio $q = 0.558 \pm 0.002$ and a secondary mass \blackpen{$M_2 = 0.87 \pm 0.01 $ \Msun.} From the spectral energy distribution and \GAIA{} parallax we find a radius $R_1 = 2.04 \pm 0.05$ \Rsun. The grazing transit profile and spectroscopic luminosity ratio indicate $R_2 = 1.04\SPSB{+0.15}{-0.10}$ \Rsun, suggesting an \blackpen{early-K spectral type.} \textcolor{black}{We show that the heartbeat feature in the {\it TESS} light curve can be explained by time-varying ellipsoidal variation, driven by the orbital eccentricity of $e = 0.3056\SPSB{+0.0024}{-0.0026}$, and relativistic beaming of the light of the primary. We find no evidence of tidally-excited oscillations. }
%This work introduces the previously-unnoticed companion of HD 181793 and presents the system as a rare and new heartbeat stellar system.
\end{abstract}


%%%%%%%%%%%%%%%%%%%%%%%%%%%%%%%%%%%%%%%%%%%%%%%%%%

%%%%%%%%%%%%%%%%% BODY OF PAPER %%%%%%%%%%%%%%%%%%

\section{Introduction}\label{sec:introduction}

The characteristic photometric variation of heartbeat stars (HBSs) which resembles a heart electrocardiogram led to their classification by \citet{Thompson2012} as a unique category of stellar binaries. These eccentric systems 
%can face drastic 
undergo strong variations in tidal forces throughout their orbits, causing \textcolor{black}{time-varying}  geometric distortion of the stellar components. This distortion is in turn observed in light curves as brightness variations with the distinct and recognisable shape that gave them their name. In the few known cases where HBSs are also found to be eclipsing, the masses and radii of the system components, \blackpen{and hence their orbital environments} can be determined with great precision \citep{Andersen1991,Southworth2004,Torres2010}.

Since this categorisation, over 1000 HBSs have been 
%identified across a plethora of studies 
\textcolor{black}{identified}
\citep{DeCat2000,Willems2002,Handler2002,Maceroni2009} and \textcolor{black}{catalogued} \citep{Thompson2012,Beck2014,Kirk2016,Kołaczek2021,Wrona2022}. Yet these stars continue to maintain great intrigue for their properties and potential to act as extreme stellar tidal laboratories in placing constraints on our theories of dynamical tides \citep{Thompson2012}, studying oscillatory eigenmodes and energy dissipation in stars \citep{Fuller2017}, or challenging current theories of orbit circularisation \citep{Thompson2012,Wrona2022}. Eclipsing binaries, particularly HBSs, also serve as probes of stellar internal structure \citep{Tkachenko2020} and of binary evolution \citep{Wrona2022}. In addition to these characteristics, 
\textcolor{black}{the periodic tidal forcing may excite internal pulsations at integer multiples of the orbital period}
% may be present in HBSs in which the star displays  
\citep{Aerts2010,Guo2020}.
\textcolor{black}{These tidally-excited oscillations (TEOs) can, in turn, serve to probe} the causes and mechanisms of these internal pulsations \citep{Cheng2020}, such as the testing of resonance locking \citep{Witte1999} and tidal dissipation as a mechanism for orbital decay, or to place additional constraints onto the systems from known pulsation criteria \citep{Lampens2021}. HBSs %allow us to explore binary interactions between stars in greater detail as the eccentricity of the orbits creates a more
\textcolor{black}{thus allow us} to uniquely investigate tidal forces and their effects on the components of these systems \citep{Kumar1995}.

\begin{figure*}
    \includegraphics[width=\textwidth]{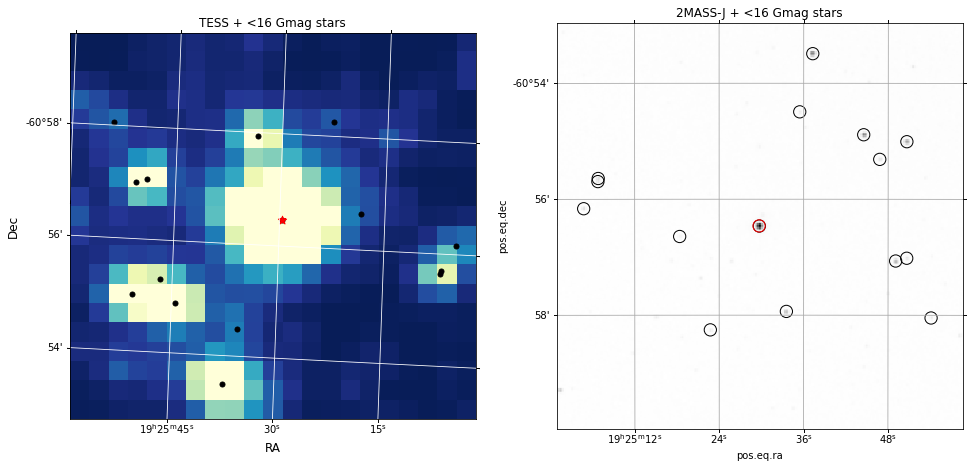}
    \caption{\redpen{(\textit{Left panel}) TESS FFI image of HD 181793 and surrounding stars and (\textit{right panel}) 2MASS map of the same area, demonstrating no significant contamination risk from surrounding stars. The 2MASS map is rotated 180$^{\circ}$ in comparison to the TESS image.}}
    \label{fig:obs:ffi}
\end{figure*}

\begin{figure}
    \includegraphics[width=\columnwidth]{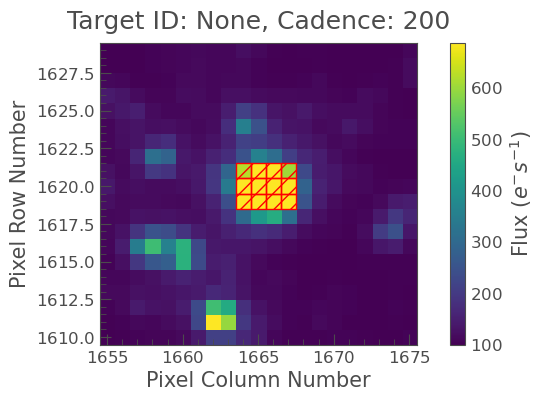}
    \caption{Selected aperture for the TESS light curves.}
    \label{fig:obs:aperture}
\end{figure}

HD 181793 was first identified by \citet{Houk1975} as a chemically peculiar (CP) Am star, a classification that is also confirmed by \citet{Egret1981} and \citet{Manfroid2009}. These stars exhibit a chemical composition showing underabundances in Ca and Sc, in conjunction with overabundances of heavier metals \citep{Conti1970}. Typically, A-type stars lie in the stellar instability strip where their neighbours in the HR diagram show brightness variations \citep{Breger1970}, but it originally appeared that Am stars did not present this same instability \citep{Cox1979}. It is now understood these stars exhibit the same rates of pulsation but at lower amplitudes than their counterparts of other stellar types \citep{Smalley2011}, but our understanding of Am stars remains in development. This stability against pulsation can lead to a better understanding of the link between the composition and structure of stars and their internal processes \citep{Paunzen2021}. Very few Am stars have been identified as HBSs with only one identified as eclipsing \citep{Kołaczek2021,Joshi2022}, thus limiting our knowledge of stellar structure that eclipsing HBSs can reveal \citep{Tkachenko2020}.

\textcolor{black}{Here we report the discovery that HD 181973 is an eclipsing HBS. 
In Section~\ref{sec:observations} we describe the discovery of eclipses in survey data from the Wide-Angle Search for Planets (WASP) \citep{wasp2006}. Light curves obtained with the Transiting Exoplanet Survey Satellite ({\it TESS}) \citep{tess2014}
%as an Object of Interest (TOI). The telescope identified a transit in the system, thus presenting the first evidence of its binary nature. At first glance, the transit also 
revealed variability characteristic of HBSs at periastron overlaid with the transit, where tidal interactions are strongest and transit probability highest. This prompted a campaign of radial velocity (RV) follow-up with the Las Cumbres Observatory (LCOGT) \citep{lcogt2013} Network of Robotic Echelle Spectrographs (NRES) instrument \citep{nres}.} 

%We then proceeded to combine radial velocity and transit models and fit these using Markov-chain Monte Carlo (MCMC) techniques to build a complete orbital model of the system, followed by spectral separation to isolate the secondary features in the spectra and determine the radial velocity of the secondary through spectral fitting. A final test was completed to create a preliminary model of the heartbeat variation in the light curve and determine its potential in future research and identify HD 181793 as a new eclipsing Am-type HBS.

%In this work, I present both these data sets as well as the complete results. Section~\ref{sec:observations} lays out the observations. 
\textcolor{black}{Section~\ref{sec:methodology} describes in detail the RV and transit models used 
to build a complete orbital model of the system, the spectral separation method to isolate the secondary features in the spectra and determine the RV of the secondary, and the heartbeat and beaming models used to fit the out-of-eclipse variations. The} results are presented and explored in Section~\ref{sec:results}, together with a discussion of the lack of tidally-excited oscillations. We  present our conclusions in Section~\ref{sec:discussion}. 
%With the distinctiveness of HD 181793, this eclipsing heartbeat binary offers a variety of avenues and intrigue for future research, and this work aims to lay the groundwork and motivation for further study of this particular system.

\begin{table}
    \centering
    \begin{tabular}{lccr}
    \hline
    \# & Obs. date & BJD & RV $\pm$ $\sigma$ (km/s)\\
    \hline
    \hline
    1 & 2022-07-01 & 2459762.8 & 45.84 $\pm$ 0.28 \\
    2 & 2022-07-02 & 2459763.7 & 40.49 $\pm$ 0.26 \\
    3 & 2022-07-03 & 2459764.7 & -0.93 $\pm$ 0.25 \\
    4 & 2022-07-04 & 2459765.8 & -36.20 $\pm$ 0.27 \\
    5 & 2022-07-05 & 2459766.8 & -46.76 $\pm$ 0.24 \\
    6 & 2022-07-06 & 2459767.8 & -45.76 $\pm$ 0.24 \\
    7 & 2022-07-08 & 2459769.7 & -31.61 $\pm$ 0.22 \\
    8 & 2022-07-11 & 2459772.6 & 13.74 $\pm$ 0.29 \\
    9 & 2022-07-12 & 2459773.6 & 34.24 $\pm$ 0.18 \\
    10 & 2022-08-10 & 2459802.7 & -43.18 $\pm$ 0.14 \\
    11 & 2022-08-24 & 2459816.6 & -19.37 $\pm$ 0.16 \\
    \hline
    \end{tabular}
    \caption{Observation details for the LCOGT spectra.}
    \label{table:obs:lco}
\end{table}

\section{Observations}\label{sec:observations}

HD 181793 was flagged as a {\it TESS} Object of Interest (TOI-1111) \citep{toicatalogue} for follow-up as a potential exoplanetary candidate due to the presence of photometric transit-like events.
The light curves are examined in-depth within this paper. Follow-up observations were made with the LCOGT NRES to provide complementary RV curves and spectral information for the stellar system. These new observations are presented in this paper.

\subsection{\textit{TESS} photometry}\label{sec:obs:photometry}

Details of the mission and operation details for the {\it TESS} telescope can be found in \citet{tess2014}. The telescope observes the sky in sectors of combined field-of-view 24$^{\circ}$ × 96$^{\circ}$ over a period of approximately 27 days each. Each \textcolor{black}{of its four 10-cm cameras} has a focal ratio \textit{f}/4, with a bandpass from 600-1000 nm. 

HD 181793 (TIC 412014494) was observed in full-frame images (FFI) obtained by {\it TESS} in sectors 13 (19 June-18 July 2019) with 30-minute cadence, and 27 (04-30 July 2020) with 2-minute cadence. \textcolor{black}{We extracted the light curves from FFI held in} the MAST database with the \texttt{TESScut} \citep{tesscut} package, and subsequently analysed the 2-minute cadence data from sector 27 with \texttt{lightkurve} \citep{lightkurve}.
We chose to use only the high-cadence Sector 27 for our light curve and heartbeat fittings, which exhibited two eclipses. 
%As seen from the FFI for the system as shown in Fig.~\ref{fig:obs:ffi} and Fig.~\ref{fig:obs:aperture}, we found no significant or concerning contamination from neighbouring stars and thus the transit was considered suitably undisturbed, and we proceed safely with the light curves considered as having negligible contamination.

Once the transits were loaded in using \texttt{lightkurve}, a box least-squares fit was conducted to determine the transit period, duration and depth. The phase-folded light curve was noted to show an unusually short duration with a V-shaped transit, the first indicators of the eccentricity of the orbit, the grazing nature of the eclipses, and the system's binary rather than exoplanetary nature. 

\begin{figure}
    \includegraphics[width=\columnwidth]{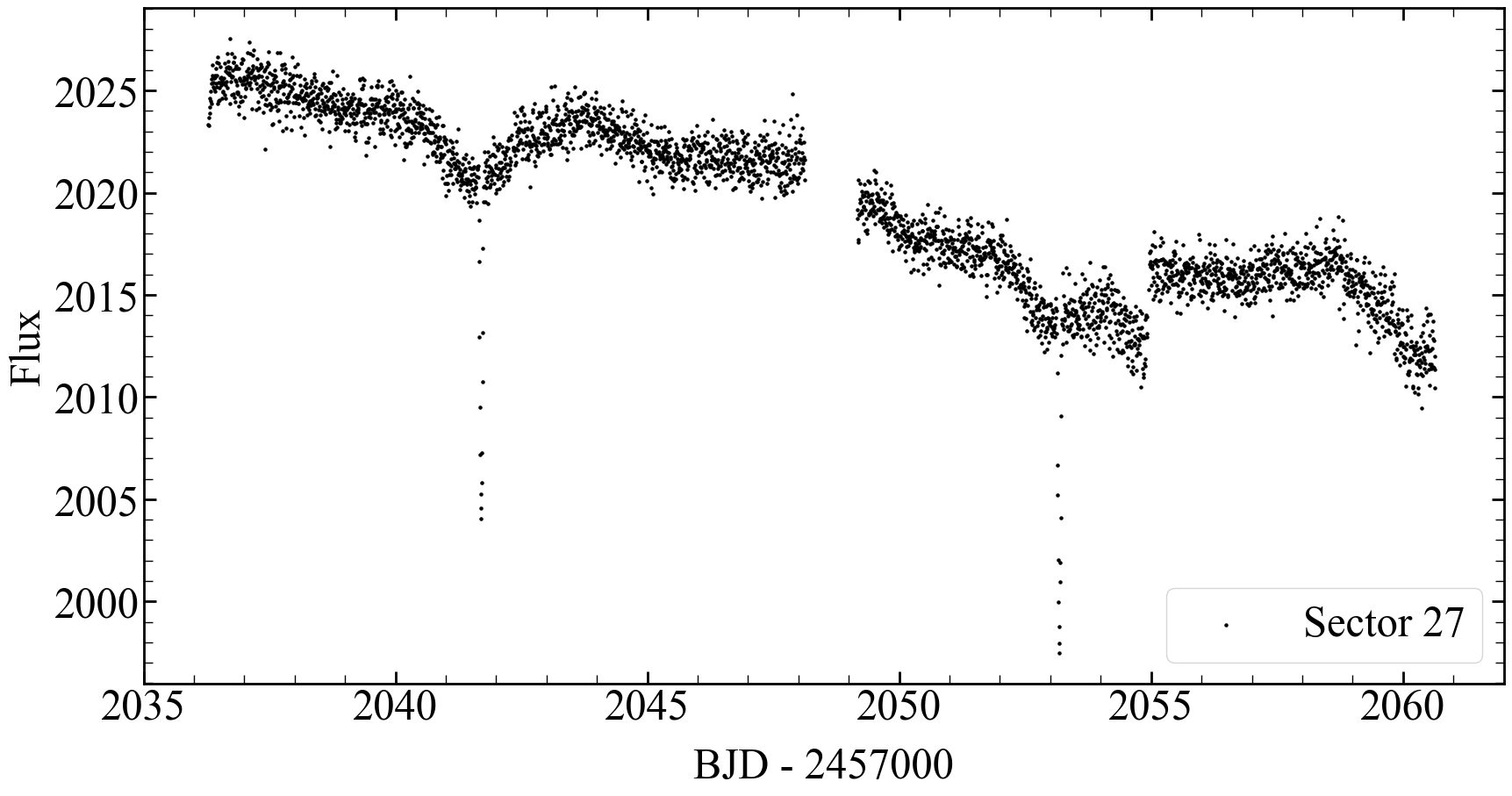}
    \caption{Raw {\it TESS} light curve of sector 27 of HD 181793. We observe a clear deep and narrow transit, around which there is a perturbation pattern that hinted at a heartbeat. We also note the momentum dump at roughly 2055 days.}
    \label{fig:obs:lc}
\end{figure}

The {\it TESS} light curve was also further processed and flattened using \texttt{lightkurve} and \texttt{PSF-SCALPELS} \citep{Wilson2022}. This employed a principal component analysis across the background, measuring the variation in flux to remove it as scattered light, followed by a linear de-trending using co-trending basis vectors and quaternions to remove the effect seen in the raw light curve (Fig.~\ref{fig:obs:lc}) caused by the spacecraft momentum dump.

\begin{figure*}
    \includegraphics[width=\textwidth]{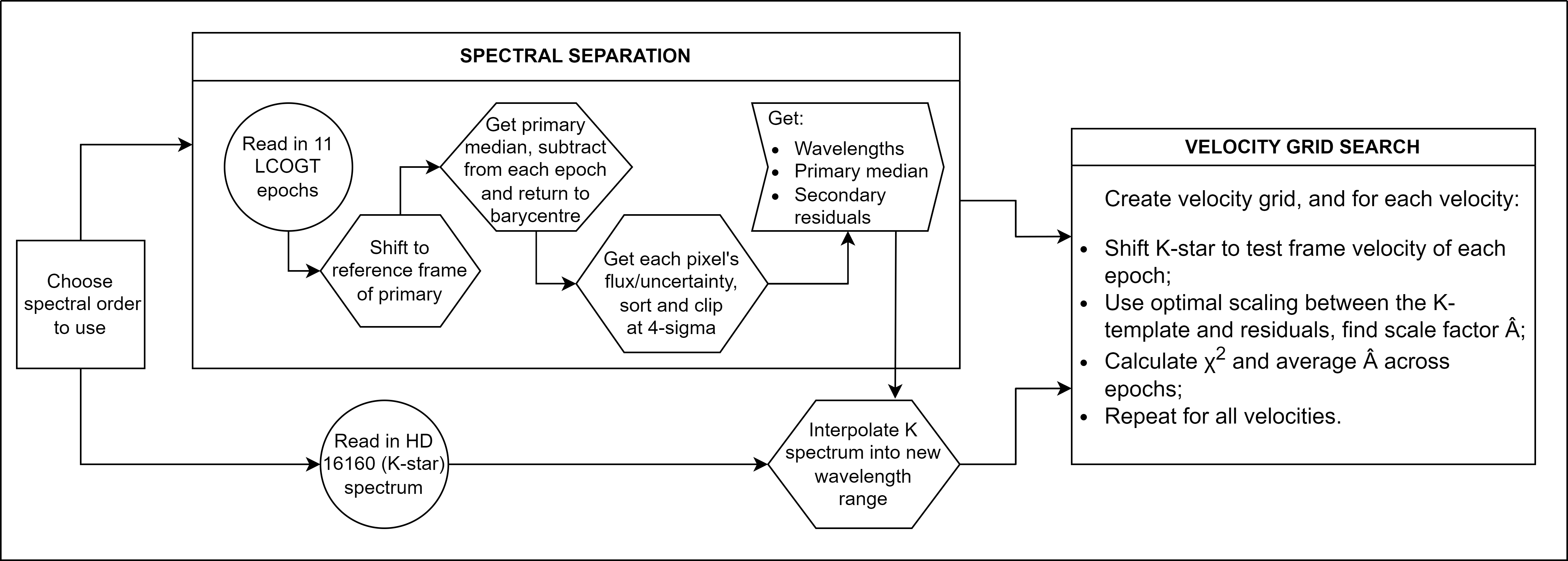}
    \caption{Flowchart presenting the algorithm used for the spectral separation. The algorithm described is that for a singular order across all epochs and velocities, and this method was then used across all orders to obtain a final result of the fitting.}
    \label{fig:met:algorithm}
\end{figure*}

\blackpen{\subsection{WASP photometry and ephemeris refinement}}

\blackpen{Six complete and five partial transits of HD 181793 were observed between 2012 July 03 and 2014 November 10 with the WASP-South instrument at Sutherland Observatory, South Africa. Like its northern counterpart SuperWASP \citep{wasp2006}, WASP-South comprised an array of eight 200-mm, \textit{f}/1.8 Canon lenses backed by 2k × 2k Peltier-cooled CCDs on a robotic mount. Each field was observed with typically 10-min cadence. A transit search triggered by its identification as TOI-1111 led to a clear detection of the 11.476-day eclipse period. We combined the historical WASP ephemeris with the {\it TESS} time of mid-eclipse to obtain a long-baseline ephemeris BJD = 2458663.0019(1) + 11.475756(1)E.} This ephemeris was then used as a prior for subsequent MCMC fitting of the light curve and orbit.\\

\subsection{LCOGT NRES follow-up}\label{sec:obs:nres}

After preliminary examination of the {\it TESS} and WASP light curves revealed this abnormally deep yet narrow transit in contrast to a relatively long period of approximately 11.5 days, follow-up observations were conducted on the NRES instrument on one of the three LCOGT 1-m telescopes located at the Cerro-Tololo Inter-American Observatory (CTIO). The resolving power of NRES is $R=53,000$, covering  the optical wavelength range (380-860 nm). A full description of the capabilities and specifications of the LCOGT is available at \citet{lcogt2013} and those for NRES at \citet{nres}. 

11 spectra were obtained between 2022 July 1 and August 24. The observation requests were timed to ensure that all orbital phases were sampled as uniformly as possible. All spectra had exposure times of roughly 2812s and a signal-to-noise ratio (SNR) greater than 30. The spectra were extracted and their RVs were measured with the BANZAI-NRES pipeline \citep{banzai2018}. The resulting RVs had a typical precision between 140 and 290 m s$^{-1}$. The log of RV measurements is given in Table~\ref{table:obs:lco}.

\section{Methodology}\label{sec:methodology}

\subsection{Orbital modelling}\label{met:orbit}

We aimed to combine both RV and transit light curve data and fit a complete model to these data. For this purpose we used the \texttt{allesfitter} package \citep{allesfitter-code,allesfitter-paper}, specifically using \texttt{ellc} \citep{ellc} and \texttt{emcee} \citep{emcee} to conduct simultaneous MCMC fitting of the RV curve and the primary eclipse profile. 
%We modelled both the primary and secondary eclipse for further confirmation of the system eccentricity. 

The primary parameters fitted by the model were the ratio of the stellar radii $R_2/R_1$; the ratio of the sum of radii to the semi-major axis $(R_1 + R_2)/a$; period $P$; time of mid-transit $T_0$; system inclination cos $i$; RV semi-amplitude of the primary $K_1$; the surface brightness ratio $J$; and the eccentricity parameters $\sqrt{e} \cos \omega$ and $\sqrt{e} \sin \omega$. Additionally, a dilution parameter for {\it TESS} was included and a quadratic limb-darkening law was chosen for the {\it TESS} data. For the derived parameters, we also included priors based on the \redpen{Gaia DR3 \citep{Gaia2016, Gaia2018, Gaia2021} radius estimate of 1.963 $\pm$ 0.041 \Rsun{} from GSP-Spec FLAME and a weighted average for temperature of \textit{T\textsubscript{eff}} = $ 7661 \pm 12$ K using the \textit{T\textsubscript{eff}} $= 7768 \pm ^{13}_{11} $ K GSP-Spec Matisse-Gauguin and \textit{T\textsubscript{eff}} = $ 7130 \pm ^{29}_{27}$ K GSP-Spec ANN values,} as well as a preliminary mass estimate of 1.57 $\pm$ 0.01 \Msun\ obtained from initial runs of the fitting and separation algorithm. %This mass will be further confirmed in the spectral separation.

The MCMC ran with 100 walkers, 1000 burn-in steps and 15000 total steps. \blackpen{From the Gelman-Rubin test, the fitting demonstrably converged to a stationary state with all chains being over 30 times the autocorrelation time}, with most parameters giving approximately Gaussian distributions. These are shown in Appendix~\ref{app:mcmc}, and the parameter priors and results are presented in Section~\ref{sec:res:orbit}.

\subsection{Spectral separation}\label{specsep}

\textcolor{black}{The secondary spectrum is not apparent in the extracted spectra, so we adopted template-matching approach to determining the RV amplitude of the secondary.}
%Once the orbital model and fit and the radial velocity parameters of the primary were obtained, the next step involved spectral separation to isolate and identify the secondary features in the LCOGT spectra. 
%Due to the features of the primary being significantly stronger than those of the secondary, and additional contamination from noise, telluric features and cosmic rays the spectra needed additional manipulation and considerations to complete the fitting algorithm. 
A flowchart of the algorithm \textcolor{black}{is} provided in Figure~\ref{fig:met:algorithm} to assist the description.

%The algorithm was built, at first, to operate for individual orders. 
\blackpen{The BANZAI-NRES pipeline returns extracted 1-D spectra of the individual echelle orders in the barycentric reference frame. In preference to merging the spectra by re-binning, we adopted an order-by-order approach. We used the Keplerian orbital solution to shift the spectral orders into the reference frame of the primary using linear interpolation.} 
%It began with a simple Doppler shifting based on the previously-obtained orbital model to align the features of the primary, and with a simple linear interpolation 
From the spectrum in each order we built a median primary spectrum and subtracted it from the individual epochs. 
%This used the \texttt{pyCHEOPS} radial velocity model \citep{pycheops}. 
%Sorting the pixels by their flux-to-uncertainty ratio showed a roughly Gaussian distribution, and 
\textcolor{black}{We performed}
a 4-$\sigma$ clip 
%was then conducted on each of the epochs 
to flag and mask outliers in the spectrum at each epoch caused by either cosmic rays
or Doppler-shifted telluric absorption features.
%only the relevant features. Outliers were then flagged as either telluric or cosmic features and masked.

The residuals were then returned to the barycentric frame to run the \textcolor{black}{template-matching analysis.} 
%From the results of the MCMC fitting, the secondary seemed to qualify as either an early K-dwarf or late G-star. 
\textcolor{black}{The RV amplitude and period of the primary's orbit, together with the initial mass estimate, suggest a secondary mass of order 0.8 M$_\odot$, and hence an early-K or late-G secondary spectral type.
\textcolor{black}{The NRES spectral standard star HD 16160 (spectral type K3V)} was identified as a %suitable K-dwarf whose 
spectral template. Its spectrum was observed on 2021 December 29  with \textcolor{black}{the NRES instrument at CTIO that had been used for the HD 181793 campaign and reduced by the BANZAI pipeline}, obtained for use through the LCOGT Science Archive.}  
%This spectrum was then used as the comparative template using which the spectral amplitude was fitted.

For each order, the template spectrum was fitted against the residuals at each epoch. 
%To do this, 
The template spectrum was shifted using \textcolor{black}{a Keplerian orbital solution scaled by a grid of RV amplitudes}, over the range $0 < K_2 < 150$\,km\,s$^{-1}$ 
%from 0 km s\textsuperscript{-1} to 150 km s\textsuperscript{-1} 
in uniform 1 km s\textsuperscript{-1} \textcolor{black}{increments.} 
%according to the orbital phase of the epoch. 
The ends of the template and the residual array were truncated to eliminate any misalignment resulting for the shifting. 

\blackpen{At each value of $K_2$, the resulting shifted template spectrum $\mathbf{p}$, which approximates the Doppler-shifted spectrum of the secondary, was scaled to fit the inverse-variance weighted residuals $\mathbf{x}$.}
%the resulting model $\mathbf{p}$ of the orbital motion of the secondary's spectrum was scaled to fit the inverse-variance weighted residuals $\mathbf{x}$.
%Once the template had been shifted to the correct orbital phase test velocity, optimal scaling against the residual epoch was implemented. 

The optimal scaling factor, 
\begin{equation}
    \hat{A} = \mathbf{p}\cdot\mathbf{\Sigma}^{-1}\cdot\mathbf{x} /
    \mathbf{p}\cdot\mathbf{\Sigma}^{-1}\cdot\mathbf{p}
    \label{eq:ahat}
\end{equation}
and badness-of-fit statistic
\begin{equation}
    \chi^2 = (\mathbf{x}-\hat{A}\mathbf{p})\cdot\mathbf{\Sigma}^{-1}\cdot(\mathbf{x}-\hat{A}\mathbf{p})^T    
    \label{eq:chisq}
\end{equation}
were both calculated and recorded at each velocity \textcolor{black}{for the time series of spectra in each {\'e}chelle order. Here $\mathbf{\Sigma}^{-1}={\rm diag}(1/\sigma^2)$, assuming independent heteroscedastic uncertainties $\sigma$ in each pixel, as computed by the BANZAI-NRES pipeline.}
The \textcolor{black}{value of $K_2$ at the} location of the $\chi^2$ minimum, as well as the maximum in the optimal scaling factor, defined the RV of the secondary. 

Inspection of the results for individual orders showed that an unambiguous signal
was detected in all orders, except those at the very bluest and reddest wavelengths. 
We confirmed that the \textcolor{black}{secondary's signal in} the long-wavelength orders was 
\textcolor{black}{compromised by many outliers arising from Doppler-shifted telluric lines and a lack of spectral features in the K-dwarf template spectrum.}
%significantly contaminated by excessive noise 
\textcolor{black}{The spectral energy distribution of the template yielded too little flux in the bluest orders to yield a reliable detection at low SNR ratios.}

\blackpen{We computed the final estimates of $K_2$ and $\chi^2$ by summing the quantities in Eqs.~\ref{eq:ahat} and \ref{eq:chisq} over all orders in the range 4300\AA{}$ < \lambda < 6500$\AA, where the signal was well-defined, instead of individually for each order.}

\begin{table*}
    \caption{Posterior distributions of the parameters as given by the joint MCMC fit of the complete orbital model. Uncertainties have been provided to \textcolor{black}{two significant figures}.} 
    %\textcolor{red}{Laura: Presentation of this table - Does it need reformatting?}}
    \label{table:res:posteriors}
    \centering
    \newcolumntype{R}{>{\raggedleft\arraybackslash}X}
    \begin{tabularx}{\textwidth}{XRR}
    \hline
    Parameter & Value & Prior\\ 
    \hline
    \hline
    \textit{Fitted parameters } & & \\
    \hline
    $R_2 / R_1$ & $0.510\SPSB{+0.074}{-0.051}$ & $\mathcal{U}(0.45,0.65)$ \\ 
    $(R_1 + R_2) / a$ & $0.1059\SPSB{+0.0051}{-0.0037}$ & $\mathcal{U}(0.08,0.12)$ \\ 
    $\cos{i}$ & $0.1344\SPSB{+0.0075}{-0.0055}$ & $\mathcal{U}(0.06,0.21)$ \\ 
    $T_{0}$ (BJD, shifted) & $2459431.8699\SPSB{+0.0040}{-0.0037}$ & $\mathcal{U}(2459052,2459054)$ \\ 
    $P$ ($\mathrm{d}$) & $11.47563 \pm 0.00012$ & $\mathcal{U}(11.4,11.6)$ \\ 
    $K_1$ ($\mathrm{km/s}$) & $47.41\SPSB{+0.13}{-0.12}$ & $\mathcal{U}(47.0,48.0)$ \\ 
    $\sqrt{e} \cos{\omega}$ & $0.3713\SPSB{+0.0039}{-0.0045}$ & $\mathcal{U}(0.2,0.4)$ \\ 
    $\sqrt{e} \sin{\omega}$ & $0.4095 \pm 0.0037$ & $\mathcal{U}(0.3,0.5)$ \\ 
    $D_\mathrm{0}$ & $0.048\SPSB{+0.059}{-0.034}$ & $\mathcal{U}(0.0,0.5)$ \\ 
    $q_{1; \mathrm{TESS}}$  & $0.059\SPSB{+0.093}{-0.044}$ & $\mathcal{U}(0.0,1.0)$ \\ 
    $q_{2; \mathrm{TESS}}$  & $0.42\SPSB{+0.38}{-0.30}$ & $\mathcal{U}(0.0,1.0)$ \\ 
    $J_{b; \mathrm{TESS}}$  & $0.066\SPSB{+0.055}{-0.047}$ & $\mathcal{U}(0.0,0.15)$ \\
    $\ln{\sigma_\mathrm{TESS}}$ ($\ln{ \mathrm{rel. flux.}})$               & $-6.4029 \pm 0.0031$ & $\mathcal{U}(-15.0,0.0)$ \\
    $\ln{\sigma_\mathrm{NRES}}$ ($\ln{ \mathrm{km/s}}$) & $-4.1\SPSB{+2.8}{-7.4}$ & $\mathcal{U}(-15.0,0.0)$\\ 
    
    \hline
    \hline

    \textit{Derived parameters} & & \\
    \hline  
    $R_1/a$                    & $0.0700 \pm 0.0011$\\ 
    $a/R_1$                     & $14.28 \pm 0.21$\\ 
    $R_2/a$                     & $0.0358\SPSB{+0.0052}{-0.0036}$\\ 
    $R_2$ ($\mathrm{R_{\odot}}$)  & $1.04\SPSB{+0.15}{-0.10}$\\ 
    $a$ ($\mathrm{R_{\odot}}$)  & $29.13 \pm 0.86$\\ 
    $a$ (AU)                    & $0.1355 \pm 0.0040$\\ 
    Mass ratio $q$ & $0.550\SPSB{+0.026}{-0.024}$ \\ 
    Companion mass $M_\mathrm{2}$ ($\mathrm{M_{\odot}}$) & $0.864\SPSB{+0.041}{-0.038}$\\
    Inclination $i$ (deg)       & $82.28\SPSB{+0.32}{-0.44}$\\ 
    Eccentricity $e$            & $0.3056\SPSB{+0.0024}{-0.0026}$\\ 
    Argument of periastron $\omega$ (deg) & $47.80\SPSB{+0.54}{-0.50}$\\ 
    Impact parameter $b$        & $1.416\SPSB{+0.080}{-0.055}$\\ 
    Transit duration $w$ (h)    & $2.534\SPSB{+0.041}{-0.036}$\\ 
    Occultation epoch $T_\mathrm{0;occ}$ & $2459439.108\SPSB{+0.017}{-0.019}$ \\
    Limb darkening; $u_\mathrm{1}$ & $0.17\SPSB{-0.13}{+0.22}$\\ 
    Limb darkening; $u_\mathrm{2}$ & $0.03\SPSB{-0.15}{+0.18}$\\ 
    \hline
    \end{tabularx}
\end{table*}

\subsection{\redpen{Spectral analysis}}

\redpen{The system parameters established in Section~\ref{sec:res:orbit} below yield $\log g=4.0\pm 0.05$ for the surface gravitational acceleration of the primary, more accurately than is possible with spectral analysis, so we adopted this value. We solved initially for the effective temperature and metallicity [M/H] of the primary by fitting a solar-abundance model to the H$\alpha$ profile in the median spectrum of the primary. We used the pySME implementation by \citet{Wehrhahn2023} of the Spectroscopy Made Easy (SME)  spectral-fitting package \citep{Valenti1996, Piskunov2017}. We used a model-atmosphere grid generated with MARCS \citep{Gustafsson2008} with a solar-abundance line list from the Vienna Atomic-line Database \citep[VALD3;][]{Piskunov1995, Ryabchikova2015}. We used a set of spectral masks spanning the wavelength range from 3922\AA\ to 5100\AA, avoiding very strong lines such as the Balmer sequence. These regions cover many of the strong lines of the species of interest for validating the primary’s Am status, as well as several hundred lines of Fe I and Fe II.}

\subsection{Heartbeat modelling}

Having established the fundamental parameters of the system, we proceeded to examine the heartbeat feature seen at periastron. %\textcolor{black}{Although the heartbeat feature is clearly present in both halves of the sector 27 light curve,} we restricted the fitting only to the first half, 
%of the Sector 27 light curve
%\textcolor{black}{the second half being contaminated by systematic errors arising from} a momentum dump. 
\textcolor{black}{Our analysis of the out-of-transit variation thus utilises}
%The  2-minute , thus leaving us with 
\blackpen{the cleanest light curve data set at our disposal from {\it TESS}, Sector 27. We do not use the WASP light curves for this fitting, since their photometric flux uncertainties are significantly greater than the amplitude of the heartbeat.}

%The model used for this is 
We used the model of ellipsoidal variation by \citet{Kumar1995} (their Eq. 44), which was originally developed to model geometric variation in eccentric binary pulsars. The basis of the model is that within eccentric orbits, the change in tidal forces can lead to the physical distortion of both components in the system, which is witnessed in the light curve as brightness variability as the stars change shape. This model has more recently been
%later commonly 
applied to a variety of HBSs \citep{Thompson2012,Wrona2022,Kołaczek2021}:

\begin{equation}
    \frac{\delta F}{F} = S \frac{1 - 3 \sin^2 i \sin^2(\nu(t) - \omega)}{(R(t)/a)^3} + C ,
\end{equation}

\begin{equation}
    R(t) = \frac{a(1 - e^2)}{1 + e \cos(\nu(t))} ,
\end{equation}

\noindent where $\delta$F/F is the fractional change in flux, $\nu$(\textit{t}) and \textit{R(t)} are respectively the true anomaly and separation of bodies at time \textit{t}, $\omega$ is the longitude of periastron, \textit{a} the semi-major axis and sin \textit{i} the inclination. The scale factor \textit{S} and zero-point \textit{C} were fitted using the least-squares minimisation package \texttt{lmfit} \citep{lmfit}. Though the 
%so-called 
Kumar model 
%retains 
\textcolor{black}{has limitations described by} \citet{Thompson2012} and \citet{Kołaczek2021}, it 
%presents a model that 
balances mathematical simplicity with a good first-order approximation. 

The secondary is relatively faint, so reflection effects are negligible and the elliptical effect is already accounted for by the \citet{Kumar1995} model. The orbital velocity amplitude of the primary is sufficient for relativistic Doppler beaming to contribute significantly to the out-of-transit variation and needs to be accounted for in the modelling. We incorporated beaming in the model using 
the formulation from the BEER algorithm \citep{Faigler2011}, applied to the primary's Keplerian  orbit model:
%were added in as a test of Doppler beaming. While , Doppler beaming was a main point of interest due to the high velocity of the primary star. We added this in with a model based off \citet{Faigler2011}'s Eq. (7) accounting for eccentric change in velocity using the \texttt{pyCHEOPS vrad} function, which was then combined with Faigler's model:

\begin{equation}
    \frac{\delta F}{F} = - 4 \alpha\textsubscript{beam} \frac{V\textsubscript{rad}}{c}.
\end{equation}

%Noting that the BEER model uses a sine curve for the radial velocity while \texttt{pyCHEOPS} uses a cosine, the velocity curve required a factor of -1 to match. This negative sign was thus accounted for in the model. 
The BEER model expects the value for FGK stars to be roughly 0.8-1.2 and we thus set the value of \textit{$\alpha$\textsubscript{beam}} to 1.0. While \textcolor{black}{the primary is an Am star}, we aimed only to test the impact of Doppler boosting on the heartbeat.

The complete model combines the \citet{Kumar1995} heartbeat model and the \citet{Faigler2011} BEER Doppler beaming effect: 

\begin{equation}
    \begin{split}
    \frac{\delta F}{F} = C+S\frac{1 - 3 \sin^2 i \sin^2(\nu(t) - \omega)}{(R(t)/a)^3} - 4 \alpha\textsubscript{beam} \frac{V\textsubscript{rad}}{c}.
    \end{split}
\end{equation}

%This yielded the complete model for the heartbeat fitting that we proceeded with using.

%%%%%%%%%%%%%%%%%%%%%%%%%%%%%%%%%%%%%%%%%%%%%%%%%%%%%
%%%------------------RESULTS----------------------%%%
%%%%%%%%%%%%%%%%%%%%%%%%%%%%%%%%%%%%%%%%%%%%%%%%%%%%%

\section{Results}\label{sec:results}

\subsection{Orbital model}\label{sec:res:orbit}

The results of the complete orbital model are displayed in Table~\ref{table:res:posteriors}, including the fit parameter priors and posteriors as well as the parameters derived by \texttt{allesfitter}. 

The \blackpen{fitted orbital inclination is} $\cos i = 0.1344\SPSB{+0.0075}{-0.0055}$, yielding an inclination 
%of approximately 
$i = 82.28\SPSB{+0.32 }{-0.44 }$ deg. The system has an eccentricity of $e = 0.3056\SPSB{+0.0024 }{-0.0026 }$ with an argument of periastron of $ \omega = 47.80\SPSB{+0.54 }{-0.50 }$ deg. The eclipse impact parameter is $b = 1.416\SPSB{+0.080}{-0.055}$, \textcolor{black}{in units of the radius of the primary}. This confirms the extreme grazing nature of the eclipse and matches the very short eclipse duration of $w = 2.534\SPSB{+0.041}{-0.036}$ hours. We further refine the ephemeris obtained from WASP, and use the time of mid-transit from {\it TESS} to obtain an orbital period of $P = 11.47578275 \pm 0.00000055$ days.

%We also present a very precise ephemeris for the system of: 
%
%\begin{equation}
%    T_{BJD} = 2459431.8701 + 11.47564 \cdot E.
%\end{equation}

\begin{figure}
    \includegraphics[width=\columnwidth]{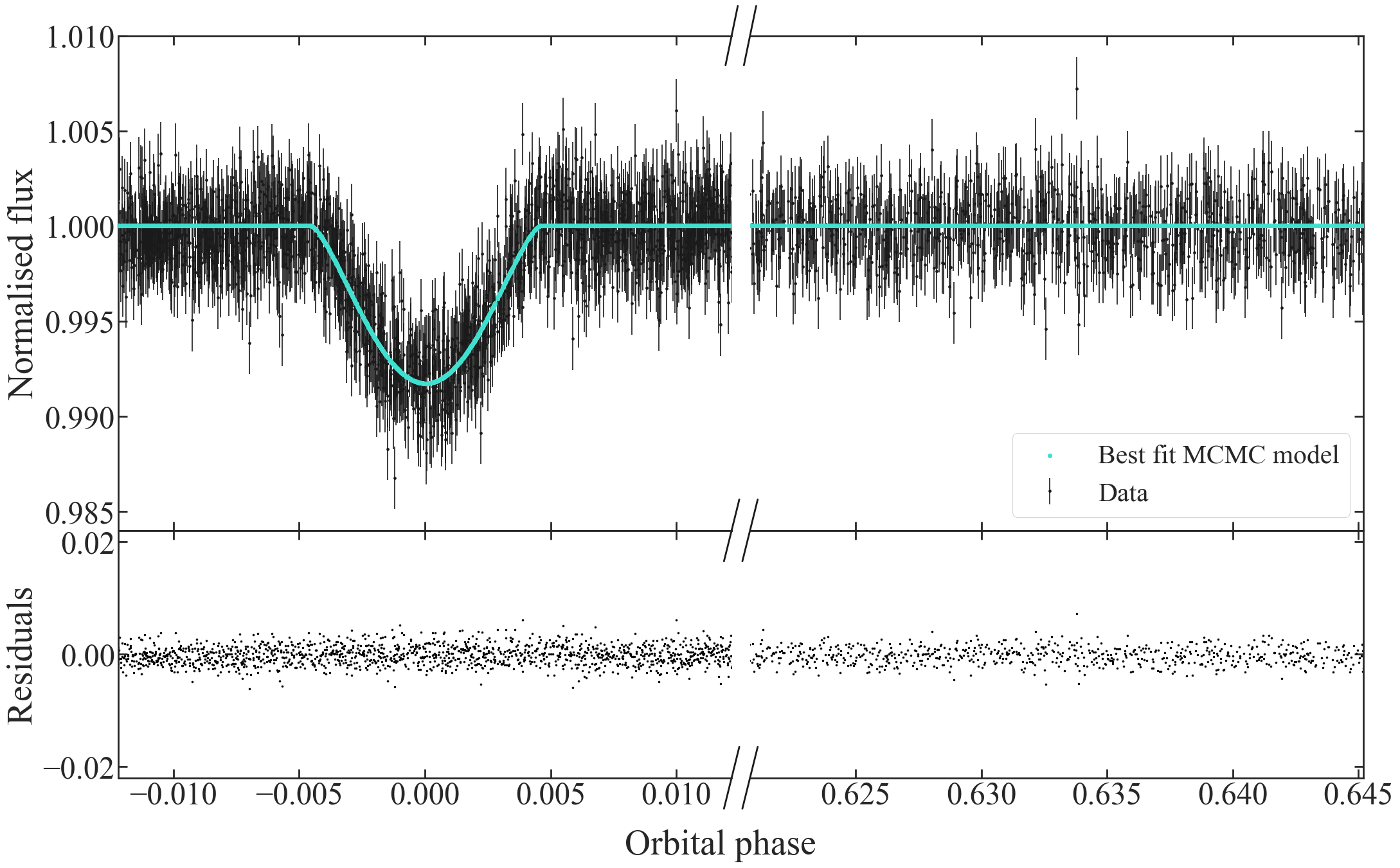}
    \caption{Results of the joint fit for the eclipse. Top panel shows the scattered data points and the MCMC best-fit model, and the bottom panel shows the residuals. \blackpen{The primary eclipse is seen centred at orbital phase zero, but no secondary eclipse is seen or expected at phase 0.633 where superior conjunction of the secondary occurs, owing to the eccentricity and inclination of the orbit.}}
    \label{fig:res:transitfit}
\end{figure}

Given the grazing primary eclipse and the eccentricity of the system, no secondary occultation was expected or detected. We show the transit model and the photometry in Fig.~\ref{fig:res:transitfit} at \blackpen{phases 0.0 and 0.633 where the primary and secondary eclipses are respectively expected, given $e=0.3056$ and $\omega=47.8\deg$}.

We record the individual stellar parameters with \redpen{the primary radius as $ 2.04 \pm 0.05 $ \Rsun}. We fitted a semi-major axis length of $a = 29.13 \pm 0.86$ \Rsun{} and a secondary radius of  $R_2 = 1.04\SPSB{+0.15}{-0.10}$ \Rsun. This would give a preliminary indication of a mid-type G-type star, with a fitted surface brightness ratio $J_b = 0.066\SPSB{+0.055}{-0.047}$ which we will later compare to the flux ratio obtained in the spectral separation. 

\begin{figure}
    \includegraphics[width=\columnwidth]{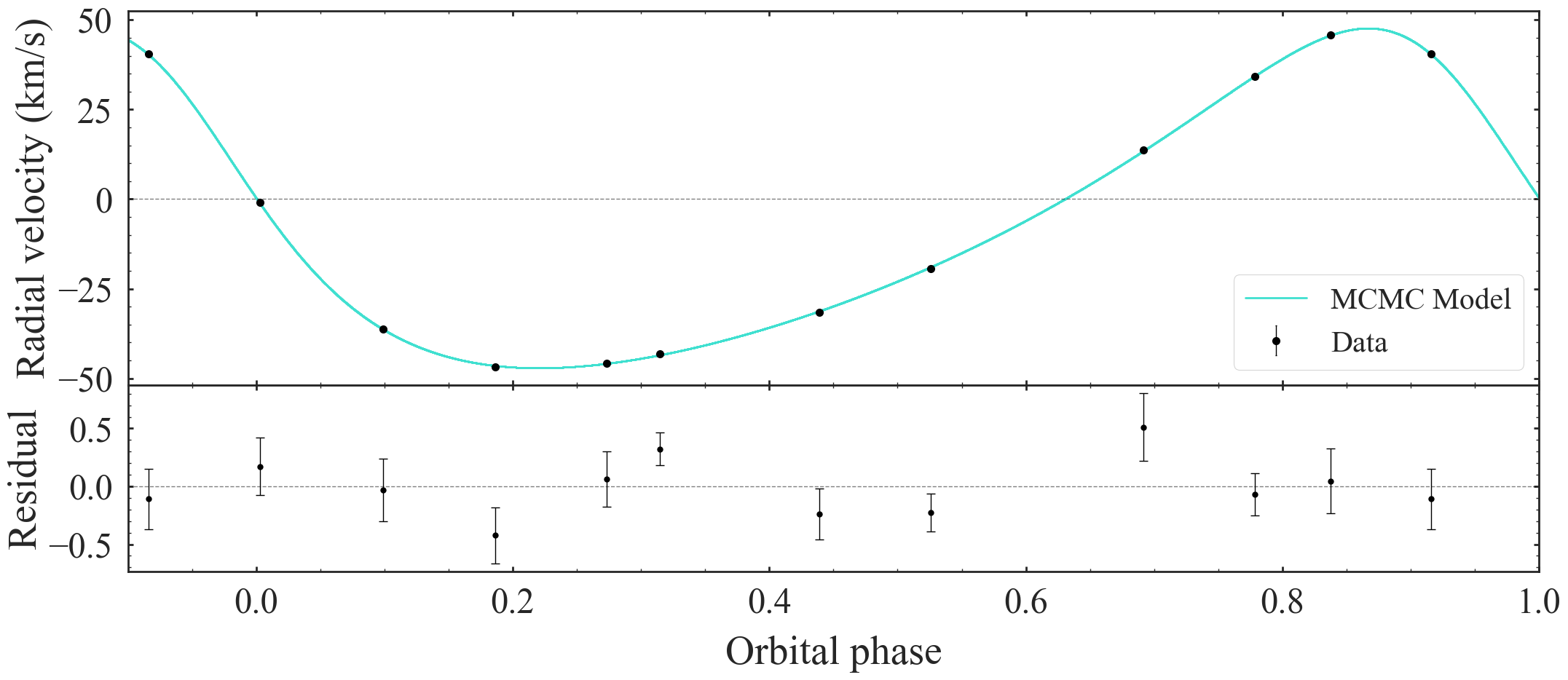}
    \caption{Results of the joint fit for the RV data. Top panel shows the data points and the best-fit MCMC model, and bottom panel shows the residuals. Note that though the data uncertainties are present in the top panel, they are smaller than the marker size and are not visible. The RV errors can be seen more clearly in the residual panel.}
    \label{fig:res:radvelfit}
\end{figure}

The RV model, displayed in Fig.~\ref{fig:res:radvelfit}, establishes the RV of the primary star as $K_1 = 47.41\SPSB{+0.13}{-0.12}$ km s\textsuperscript{-1} and gives an initial mass ratio estimate of $q = 0.550\SPSB{+0.026}{-0.024}$.

%We also show in Table~\ref{table:res:dispersion} the dispersion of radial velocities . They 
The residuals between the model values and the measured radial velocities show a root-mean-square velocity dispersion of 0.253 km s\textsuperscript{-1}, which is comparable to the uncertainties estimated by the BANZAI-NRES pipeline. There is no evidence of any significant additional sources of systematic error.

%\begin{table}
%    \centering
%    \begin{tabular}{lcr}
%    \hline
%    Phase & Measured RV & Model RV\\
%    \hline
%    \hline
%    -0.16 & 45.84 & 45.80 \\
%    -0.08 & 40.49 & 40.58 \\
%    0.00 & -0.93 & -1.09 \\
%    0.10 & -36.20 & -36.14 \\
%    0.19 & -46.76 & -46.32 \\
%    0.27 & -45.76 & -45.82 \\
%    0.44 & -31.61 & -31.39 \\
%    0.69 & 13.74 & 13.22 \\
%    -0.22 & 34.24 & 34.32 \\
%    0.31 & -43.18 & -43.51 \\
%    0.53 & -19.37 & -19.16 \\
%    \hline
%    \end{tabular}
%    \caption{A comparison table showing the measured and model RV values at each orbital phase.}
%    \label{table:res:dispersion}
%\end{table}

%We proceed with the spectral separation, adopting this orbital model for the subtraction of the primary features from the spectra. 

\subsection{Secondary radial velocity}

\begin{figure}
    \includegraphics[width=\columnwidth]{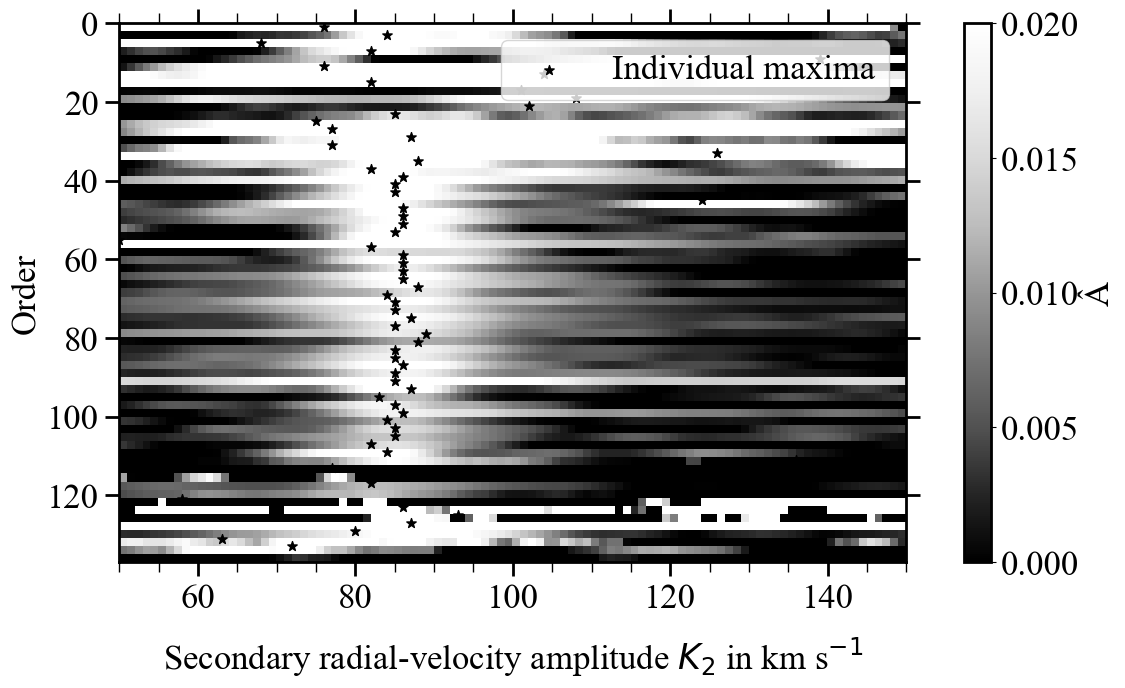}
    \caption{\textcolor{black}{Greyscale representation of the fitted flux scale factor $\hat{A}$ with the RV amplitude of the secondary.} In particular, only the central orders \textcolor{black}{numbered from} 40 to 110 show \textcolor{black}{clearly-defined} peaks.
    %, while the both ends of the grid show further evidence for accounting only for wavelengths between 4000-7000 Angstrom. 
    \textcolor{black}{In the orders with clear detections, the secondary's RV amplitude is tightly distributed around 85 km s\textsuperscript{-1}.}}
    \label{fig:res:grid_a}
\end{figure}

\begin{figure}
    \includegraphics[width=\columnwidth]{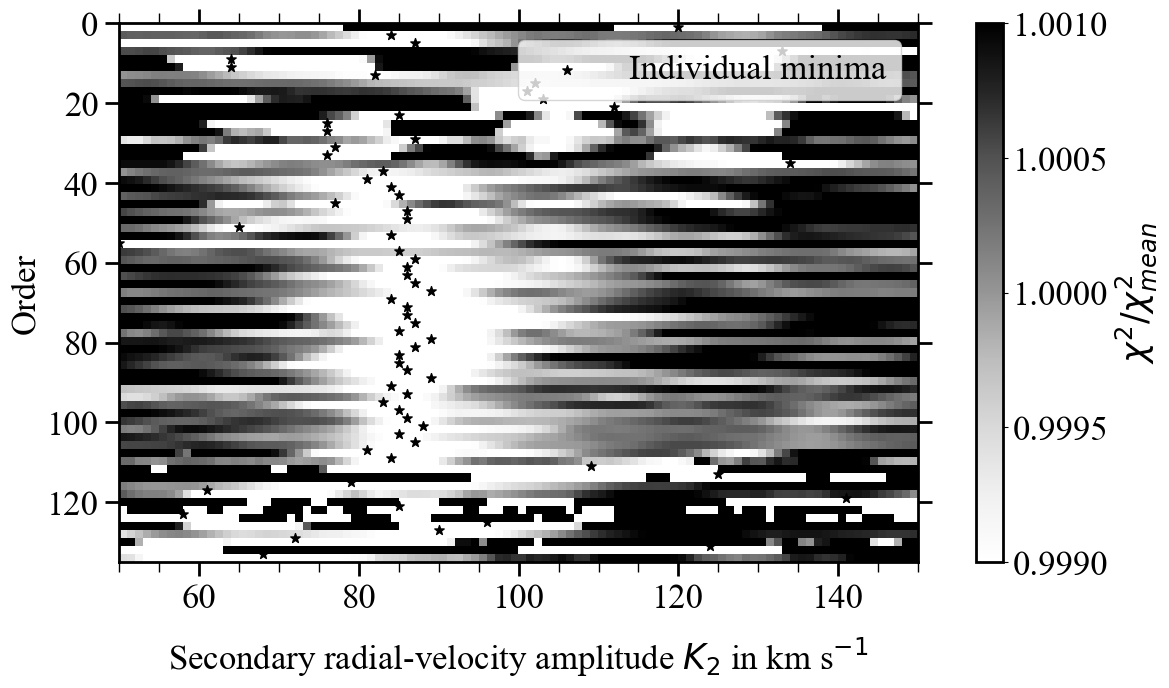}
    \caption{\textcolor{black}{Greyscale representation of} the per-order values of $\chi^2$, normalised to the mean $\chi^2$ of each order.
    %- this allowed for better visualisation due to some orders giving results of drastically higher magnitudes and thus skewing the scaling. 
    \textcolor{black}{As seen in Fig.~\ref{fig:res:grid_a}, the secondary spectrum is detected clearly in spectral orders 40 through 100, showing a tight distribution}
    %We see a similar result with the $\chi^2$ as with the optimal scaling factor, and once more reject orders outwith the 4000-7000 Angstrom range and see a general spread 
    of values around 85 km s\textsuperscript{-1}.}
    \label{fig:res:grid_c}
\end{figure}

Following the subtraction of the primary, \textcolor{black}{masking of} outlier features and the fitting of the K-star spectrum across all orders, we see a distinct localisation of the maxima in the optimal \textcolor{black}{flux-scaling} factor $\hat{A}$ as shown in Fig.~\ref{fig:res:grid_a} and a similar distribution for the order-wise minima of $\chi^2$ shown in Fig.~\ref{fig:res:grid_c}. These also \textcolor{black}{illustrate the lack of a clear detection of the secondary in the bluest and reddest orders, as noted previously.}
%present a further confirmation of the limitations placed by the long and short wavelength orders as seen by the random distribution of the maxima and minima in these grids. 

\begin{figure}
    \includegraphics[width=\columnwidth]{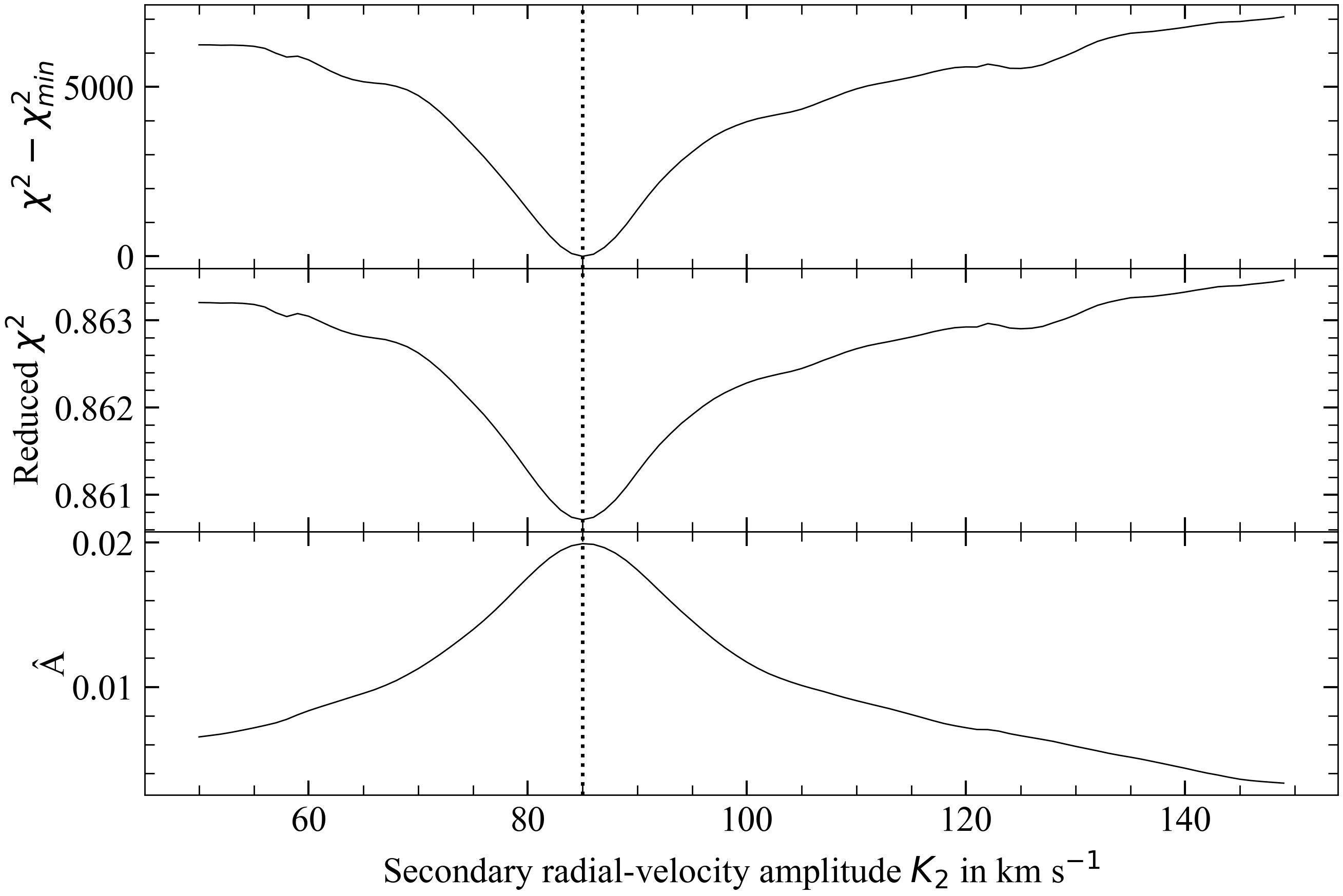}
    \caption{Chi-squared and optimal scaling factor summed across wavelengths 4300-6500 Angstrom, showing a distinct peak and minimum at the \textcolor{black}{inferred location of the best-fitting RV amplitude of the secondary.}}
    \label{fig:res:chi-squared}
\end{figure}

\begin{figure*}
    \includegraphics[width=\textwidth]{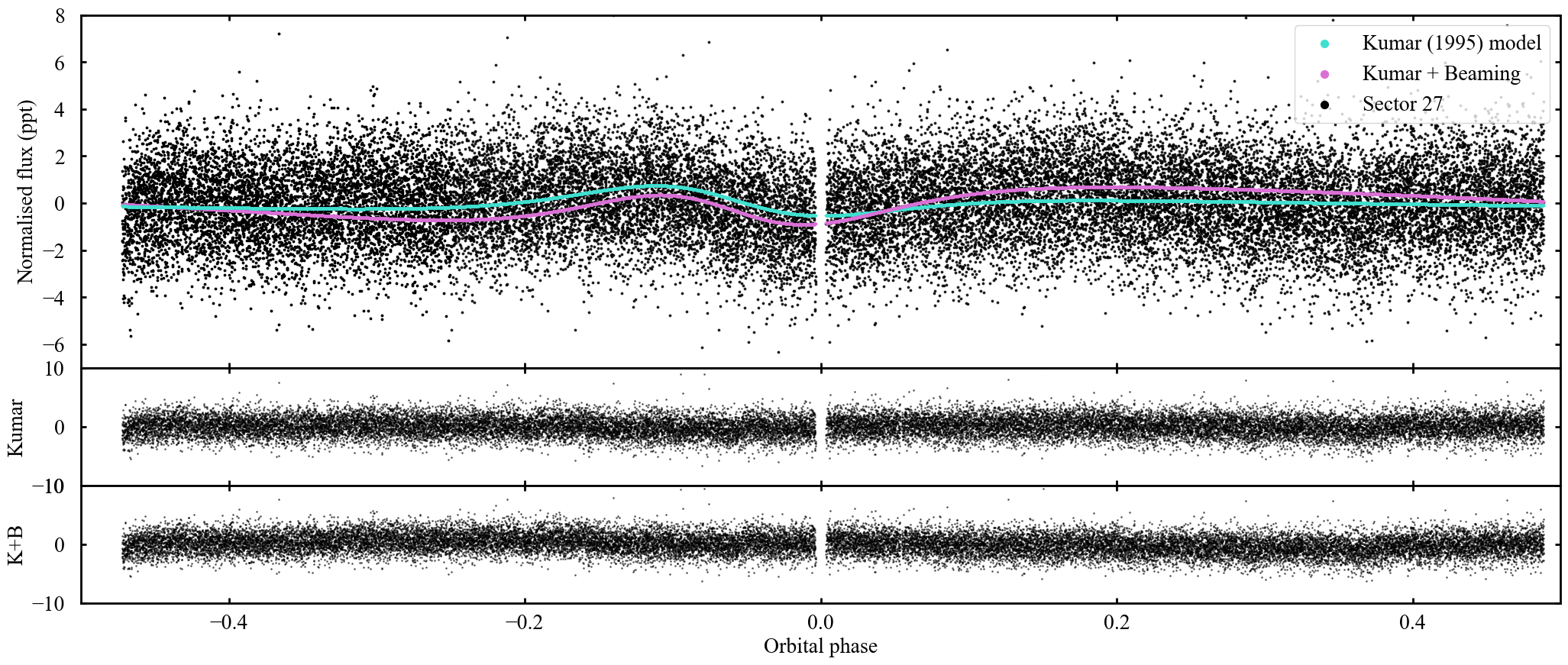}
    \caption{Data for the first half of the {\it TESS} sector 27 light curve compared with the fitted heartbeat modelled following the description in \citet{Kumar1995}, \textcolor{black}{and a full model} including the Doppler beaming of the primary with the eclipse removed from the modelling. We note that both models show roughly flat residuals and model the pattern of the heartbeat well. We conclude that the \textcolor{black}{dominant} cause of the light curve variation is the ellipsoidal variation \textcolor{black}{of the primary} from the heartbeat.}
    \label{fig:res:heartbeat}
\end{figure*}

\textcolor{black}{Conducting the scaling-factor summation over all orders between 4300\,\AA\ and 6500\,\AA}, we obtained the average optimal scaling factor and the summed $\chi^2$, \textcolor{black}{whose variations with $K_2$ are} displayed in Fig.~\ref{fig:res:chi-squared}. We see a \textcolor{black}{well-defined} $\chi^2$ minimum and a maximum in the optimal scaling factor.

\textcolor{black}{To estimate the uncertainty in $K_2$ we repeated the grid search
over a restricted range 70 km\,s$^{-1} < K_2 < 100$\,km\,s$^{-1}$, }
%Having obtained this optimal velocity, we narrowed the range and increased the resolution of the velocity grid search to range from 70 to 100 km s\textsuperscript{-1} 
at 0.05 km s\textsuperscript{-1} intervals. We fitted a quadratic function around the location of the minimum in $\chi^2$ to determine the curvature and hence the 
$\Delta \chi^2 = 1$ error bar. \textcolor{black}{This}
%With a quadratic function fitted to the minimum location, this then ultimately 
yielded an optimal RV semi-amplitude $K_2 = 84.95\SPSB{+0.12}{-0.09}$ km s\textsuperscript{-1} \textcolor{black}{for the secondary's orbit}. This fitting is \textcolor{black}{shown in more detail} in Appendix~\ref{app:spectralfit}.

Combining this measurement with the RV semi-amplitude $K_1 = 47.41 \pm 0.13$ km s\textsuperscript{-1} of the primary, we obtain a confirmed mass ratio $q = 0.558 \pm 0.002$ for the system. This is slightly larger than the result we found from the MCMC fitting ($q = 0.550\SPSB{+0.026}{-0.024}$), but is in agreement within 1-$\sigma$. With the lack of precision in the original estimate of the primary's mass, it was expected that the MCMC's results for the mass would be less reliable than that obtained from the spectral separation. Kepler's equations for eccentric orbits

\blackpen{
\begin{equation}
    a = \frac{P}{2 \pi }\frac{\sqrt{1 - e^2}}{\sin i}(K_1 + K_2),
\end{equation}
}
and
\begin{equation}
    M_{tot} = \frac{a^3}{G} \left(\frac{2 \pi}{P}\right)^2,
\end{equation}

yield a semi-major axis \blackpen{$a = 28.8 \pm 0.05$ \Rsun{}}, a value that is more precise than the MCMC fit result. Both values are consistent with each other within the \blackpen{MCMC} uncertainty bounds. The total system mass is found to be \blackpen{$M_{\rm tot} = 2.44 \pm 0.02$ \Msun{}. The primary and secondary masses are $M_1 = 1.57 \pm 0.01$ \Msun{} and $M_2 = 0.87 \pm 0.01$ \Msun{}} respectively.

\textcolor{black}{It is also noted that the peak value of the optimal scaling factor lies at 0.020, giving the spectral flux ratio and thus consistent with a surface brightness ratio $J = 0.066\SPSB{+0.055}{-0.047}$ as given by the MCMC fitting related by the equation:}

\begin{equation}
    f_R = \left( \frac{R_2}{R_1} \right)^2 \cdot J,
\end{equation}

\textcolor{black}{which returns a flux ratio $f_R = 0.017 \pm 0.014$. These values agree with one another (though we note the large uncertainty in the flux ratio caused by the large uncertainty in the MCMC surface brightness ratio, and this is likely caused by the significant difference in detectability of the two stars).}

\redpen{\subsection{Spectral type}}

\redpen{
Having established $T_{\rm eff} = 7200\pm 200$K and adopting [M/H] = $0.0\pm 0.2$, we solved for individual abundances of the atomic species Ca, Sc, Ti, Cr, Fe, Zn, Sr, Zr and Ba. We found logarithmic departures from solar abundances relative to hydrogen, displayed in Table~\ref{table:res:abundance}
for $7000{\rm K} < T_{eff} < 7400{\rm K}$ and $-0.2 < {\rm [M/H]} < 0.2$. The pattern of abundances seen in HD181793 is typical of that seen in recent abundance studies of chemically-peculiar Am stars, \citep[e.g.][]{Catanzaro2019, Romanovskaya2023}, with significant depletion of Ca and Sc relative to the Sun, and over-abundances of the heavier elements, particularly Ba. This supports the earlier identifications of HD 181793 as a CP Am star.}

\begin{table}
    \centering
    \begin{tabular}{ c  c }
    \hline
    \redpen{Abundance} & 
    \redpen{Departure ($\log_{10}$) }\\
        \hline
        \hline
        
        \redpen{$[$M/H$] $} & \redpen{ $0.0 \pm 0.2$} \\
        \redpen{$[$Ca/H$]$} & \redpen{$-0.3 \pm 0.2$} \\
        \redpen{$[$Sc/H$]$} & \redpen{$-1.7 \pm 0.2$} \\
        \redpen{$[$Ti/H$]$} & \redpen{$-0.5 \pm 0.2$} \\
        \redpen{$[$Cr/H$]$} & \redpen{$0.0 \pm 0.2$ }\\
        \redpen{$[$Fe/H$]$} & \redpen{$-0.1 \pm 0.2$} \\
        \redpen{$[$Zn/H$]$} & \redpen{$0.4 \pm 0.2$ }\\
        \redpen{$[$Sr/H$]$} & \redpen{$0.6 \pm 0.25$} \\
        \redpen{$[$Zr/H$]$} & \redpen{$0.71 \pm 0.2$} \\
        \redpen{$[$Ba/H$]$} & \redpen{$0.60 \pm 0.15$} \\
    \hline
    \end{tabular}
    
    \caption{\redpen{Measured logarithmic departures from solar abundances for HD 181973 from spectral analysis, for $7000{\rm K} < T_{\rm eff} < 7400{\rm K}$ and $-0.2 < {\rm [M/H]} < 0.2$} }
    \label{table:res:abundance}
\end{table}

\subsection{Heartbeat}

A comparison of the \textcolor{black}{out-of-eclipse} light-curve variation against the fitted heartbeat model is shown in Fig~\ref{fig:res:heartbeat}. \textcolor{black}{The geometric variation in the stellar shape of the primary accounts correctly for the shape and amplitude of the light-curve variation. Doppler beaming contributes a modest additional effect.} We therefore identify the ellipsoidal variation as the primary and main component of variation in the light curve. This establishes HD 181793 as a system with a heartbeat - a rare, eclipsing Am-type HBS. We see no evidence of additional tidally-excited oscillations in the light curve.

\section{Discussion and conclusions}\label{sec:discussion}

We have established precise values for the physical parameters of both the primary and secondary stars.
Our results indicate that HD 181793 as an eccentric binary stellar system with an Am-type primary star and a likely early K-type secondary star in a grazing eclipse. 

We find the form and amplitude of the out-of-transit "heartbeat" variation to be consistent with time-varying tidal distortion of the primary, making it one of the rare Am-type heartbeat systems and only the second discovered thus far in a confirmed eclipsing binary despite the existence of over 1000 identified HBSs. The other such system discovered, SW CMa, was identified as an Am-heartbeat eclipsing binary by \citet{Kołaczek2021}. It comprises a pair of Am stars with a mass ratio of $q = 0.940 \pm 0.010$ and a period of $P = 10.091988 \pm 0.000005$ days. With this high mass ratio, SW CMa differs greatly from HD 181793. We present HD 181793 as a highly valuable resource for future research, since the secondary star contributes little to the system's luminosity, allowing the primary's variability to be studied in isolation. The difference in companion spectral type between the two systems could also provide interesting insight into any dynamical effects. It is also a particularly bright system with a V-band magnitude of 9.63 \citep{Tycho}, and thus an easy target for follow-up studies. 

The relationship of the heartbeat, periastron, and transit can also yield insight into the ways that the stars interact at closest approach in such eccentric orbits. Further and more precise modelling of the heartbeat can also produce more insight into the tidal interactions and the components contributing to the heartbeat signal and its shape. In particular, the \textcolor{black}{form of the out-of-transit variation} is still not entirely matched by the shape of the model. \blackpen{Future {\it TESS} observations will characterise the form of the heartbeat with improved precision. The precise physical parameters we have established for the two stars have strong potential to inform more sophisticated dynamical models of the tidally-driven changes in the shape of the primary.}

%As I briefly mentioned in Section~\ref{sec:introduction}, one of the features often observed on heartbeat stars are TEOs \citep{Fuller2017}. Where t
The heartbeat feature in itself is a purely geometric deformation of the star. Tidally-excited oscillations \citep{Fuller2017} are an internal pulsation of the star. As mentioned before, Am-stars appear to have observed stability against structural pulsations. Though demonstrated to pulsate at similar frequencies to other stellar types
%the same rates of frequency as their other companions 
in the stellar instability strip, Am stars generally show lower amplitudes of pulsation
%.still show resistance against pulsation in the form of lower amplitude of pulsations 
\citep{Smalley2011}. The mechanisms that drive these Am star pulsations
%and their mechanism do not appear 
are not as-yet well-understood \citep{Smalley2017}.
%, and thus what I present here may add another star to the batch of experimental targets. 
\textcolor{black}{Among the handful of Am stars that have been discovered in heartbeat binaries}, only one does not exhibit TEOs \citep{Joshi2022}. The apparent absence of such oscillations in HD 181793, as they are usually distinctly visible in the light curves, has the potential to inform and challenge theories of pulsations driven by periodic tidal forcing in an eccentric orbit, helping us to 
%also does not exhibit -- establishing in future research whether HD 181793 does or does not display these same internal pulsations as driven by the external factor of an eccentric orbit can help us 
further understand the mechanism that drives pulsations within Am stars, and obtain new insights into the structure and composition of these chemically-peculiar stars. The second of its kind, this system has allowed us to determine the masses, radii, and orbital configuration of the system, providing a comprehensive image of the binary's fundamental parameters. These can be used to model this binary more precisely, probe stellar structure and evolution, and future observations can reveal if systems akin to HD 181793 exist. What has remained outside the scope of this particular study opens a new gate and adds another peculiar system to the growing catalogue of heartbeat stars.

\section*{Acknowledgements}
%%% 
%%%TESS
This paper includes data collected by the {\it TESS} mission. Funding for the {\it TESS} mission is provided by the NASA's Science Mission Directorate. 
%%%WASP
This paper makes use of data from the first public release of the WASP data (Butters et al. 2010) as provided by the WASP consortium and services at the NASA Exoplanet Archive, which is operated by the California Institute of Technology, under contract with the National Aeronautics and Space Administration under the Exoplanet Exploration Program. 
%%%LCOGT
This work makes use of observations from the Las Cumbres Observatory global telescope network Network of Robotic Echelle Spectrographs 1 m instrument in Cerro-Tololo Inter-American Observatory. 
%%%GAIA
This work has made use of data from the European Space Agency (ESA) mission {\it Gaia} (\texttt{\url{https://www.cosmos.esa.int/gaia}}), processed by the {\it Gaia} Data Processing and Analysis Consortium (DPAC, \texttt{\url{https://www.cosmos.esa.int/web/gaia/dpac/consortium}}). Funding for the DPAC has been provided by national institutions, in particular the institutions
participating in the {\it Gaia} Multilateral Agreement. ACC and TGW acknowledge support from STFC consolidated grant numbers ST/R000824/1 and ST/V000861/1.
%%%VALD
This work has made use of the VALD database, operated at Uppsala University, the Institute of Astronomy RAS in Moscow, and the University of Vienna.
%%%PYCHEOPS
The \texttt{pyCHEOPS} Python module is an open-source software package \citep{pycheops}.
%%%LIGHTKURVE
This research made use of Lightkurve, a Python package for {\it Kepler} and {\it TESS} data analysis \citep{lightkurve}.
%%%TESSGAIAIRFM
This work has made use of the TESSGaiaIRFM code that is currently being developed by the ESA CHEOPS Data Analysis Technical Support Team.
%%%VIZIER
This research has made use of the VizieR catalogue access tool, CDS, Strasbourg, France.
%%%ALLESFITTER
This work has made use of \texttt{allesfitter} \citep{allesfitter-code,allesfitter-paper} and its affiliated packages cited below. 
%%%EMCEE
This work has made use of \texttt{emcee}, a free Python software made available under the MIT License \citep{emcee}.
%%%ELLC
This work has made use of \texttt{ellc} \citep{ellc}, a fast and flexible light curve model for binary star systems.
%%%python/gen
This work has also made use of the following software and packages: \texttt{python} \citep{python}, \texttt{numpy} \citep{numpy}, \texttt{matplotlib} \citep{matplotlib}, \texttt{scipy} \citep{scipy}, \texttt{tqdm} \citep{tqdm}, \texttt{seaborn} \citep{seaborn}, \texttt{corner} \citep{corner}. 
In order to meet institutional and research funder open access requirements, any accepted manuscript arising shall be open access under a Creative Commons Attribution (CC BY) reuse licence with zero embargoes.

%%%%%%%%%%%%%%%%%%%%%%%%%%%%%%%%%%%%%%%%%%%%%%%%%%
\section*{Data Availability}

The {\it TESS} light curves are publicly available from MAST. The LCOGT NRES spectra are available upon request from the author.

All code was built from the publicly-available packages acknowledged above.



\bibliographystyle{mnras}
\bibliography{example} % if your bibtex file is called example.bib

%\begin{thebibliography}{99}
%\end{thebibliography}

%%%%%%%%%%%%%%%%%%%%%%%%%%%%%%%%%%%%%%%%%%%%%%%%%%

%%%%%%%%%%%%%%%%% APPENDICES %%%%%%%%%%%%%%%%%%%%%

\clearpage

\appendix

\section{MCMC results}\label{app:mcmc}
\begin{figure*}
    \includegraphics[width=\textwidth]{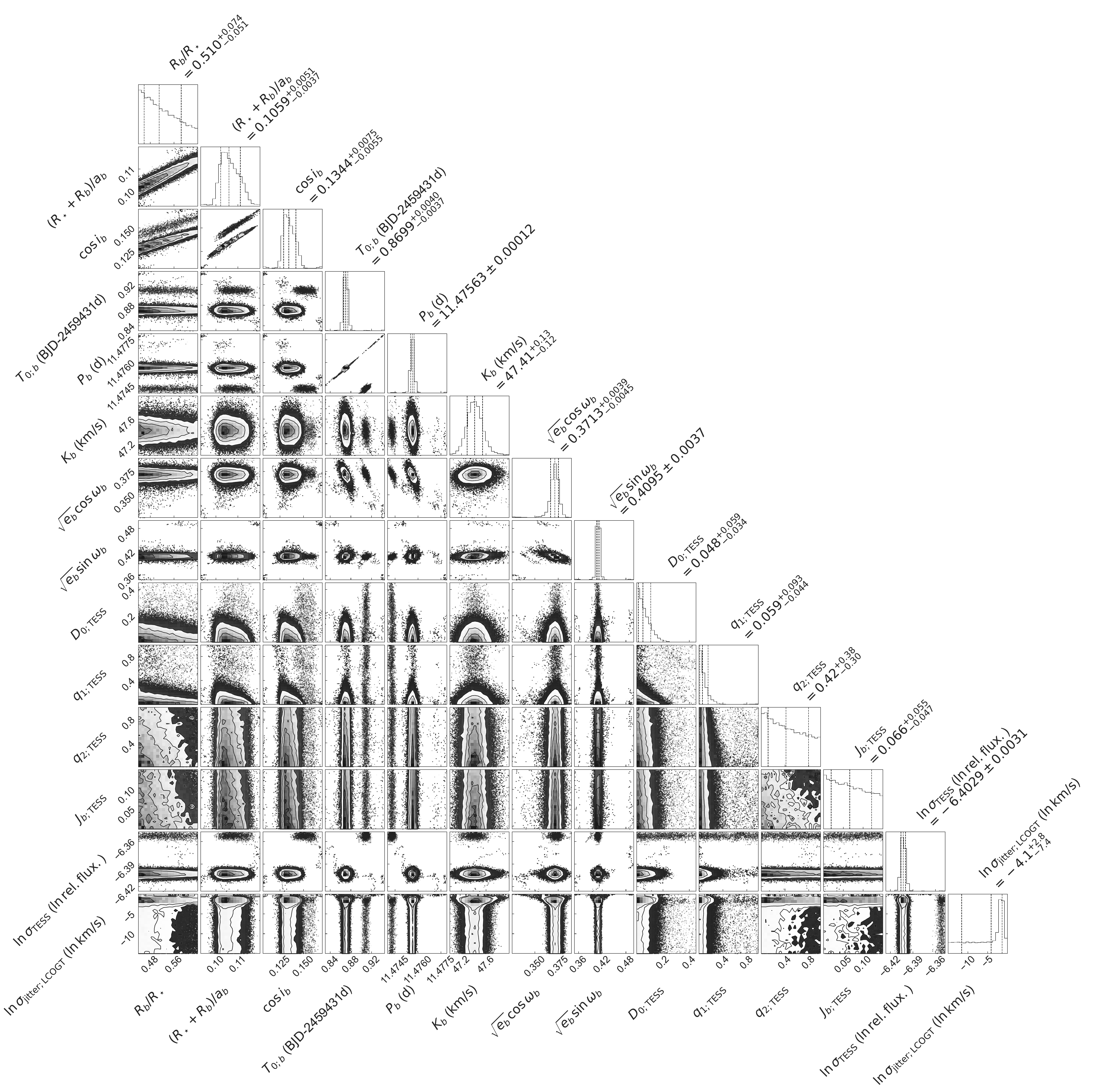}
    \caption{MCMC corner plot for the parameters fitted by allesfitter. Showing, from top to bottom, the radial ratio of the objects, the ratio of the sum of radii to the semi-major axis, inclination, epoch of mid-transit, orbital period, RV semi-amplitude, both eccentricity factors, {\it TESS} dilution factor, both host limb darkening factors, surface brightness ratio, and both error parameters.}
    \label{app:mcmc:corner}
\end{figure*}

\clearpage

\begin{figure*}
    \includegraphics[width=\textwidth]{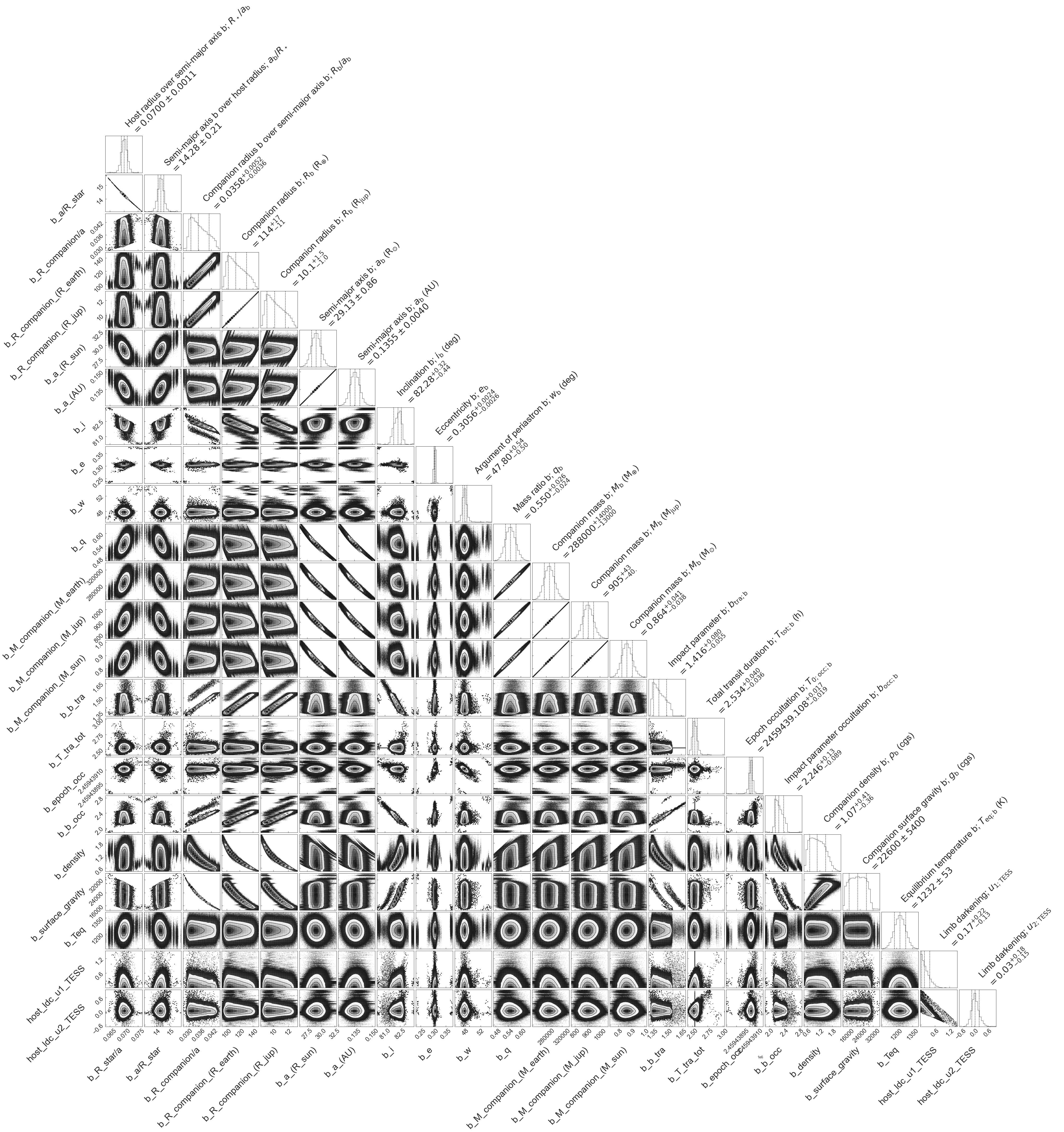}
    \caption{MCMC corner plot for the derived parameters as determined by allesfitter. Some of these parameters have not been accounted for as their values was not fitted, or remained zero, and all that were explored are shown in Table~\ref{table:res:posteriors}. Most of the parameters display Gaussian distributions, but it is likely that the non-Gaussian distributions are due to the limits imposed onto the system caused by the highly-grazing feature and eccentricity of the system. Each parameter is described within the plot, above the column showing parameter relation distributions.}
    \label{app:mcmc:derived}
\end{figure*}

\clearpage

\begin{figure*}
    \begin{subfigure}[b]{0.3\textwidth}
         \centering
         \includegraphics[width=\textwidth]{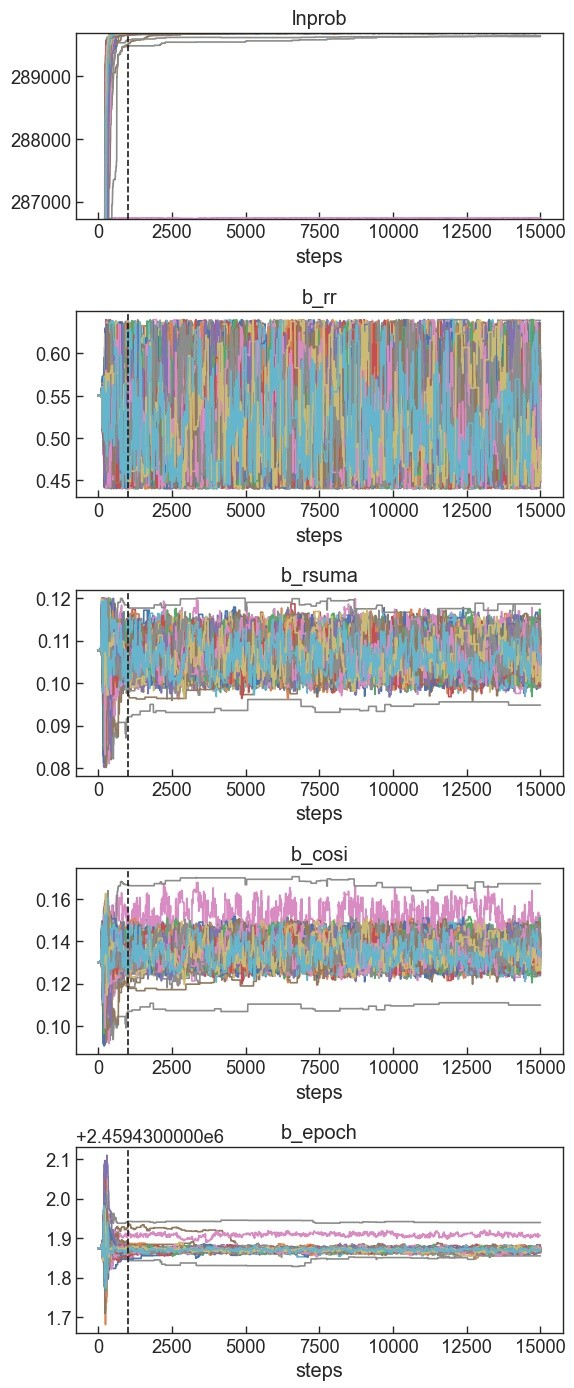}
     \end{subfigure}
     \begin{subfigure}[b]{0.3\textwidth}
         \centering
         \includegraphics[width=\textwidth]{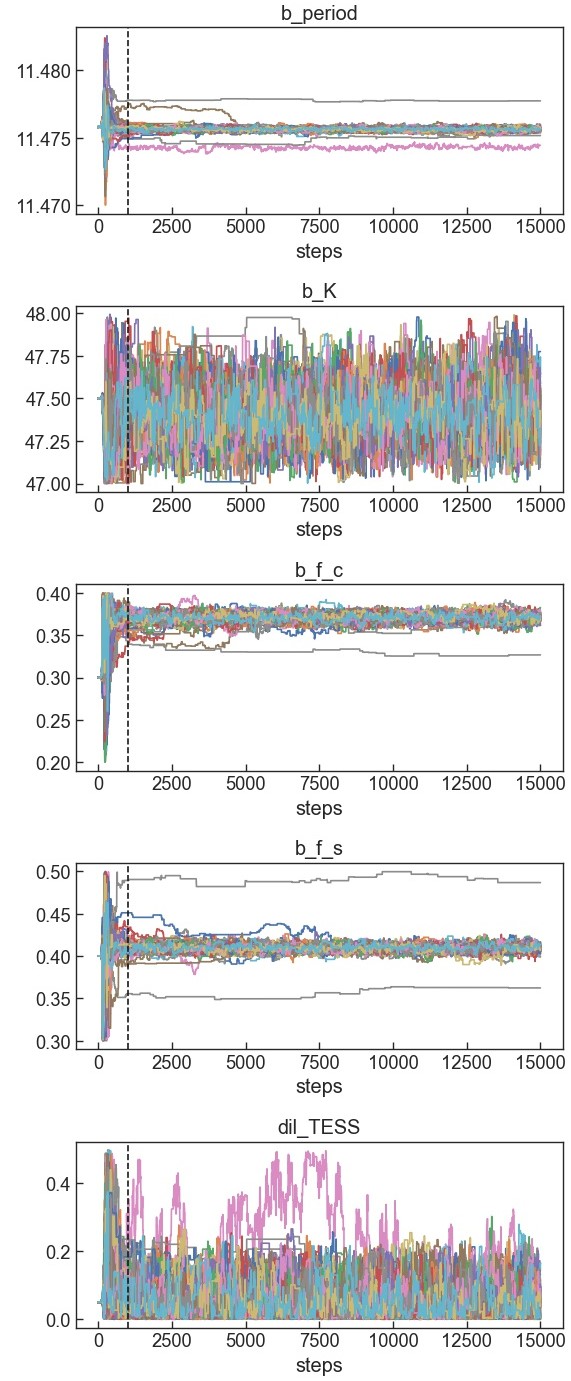}
     \end{subfigure}
     \begin{subfigure}[b]{0.3\textwidth}
         \centering
         \includegraphics[width=\textwidth]{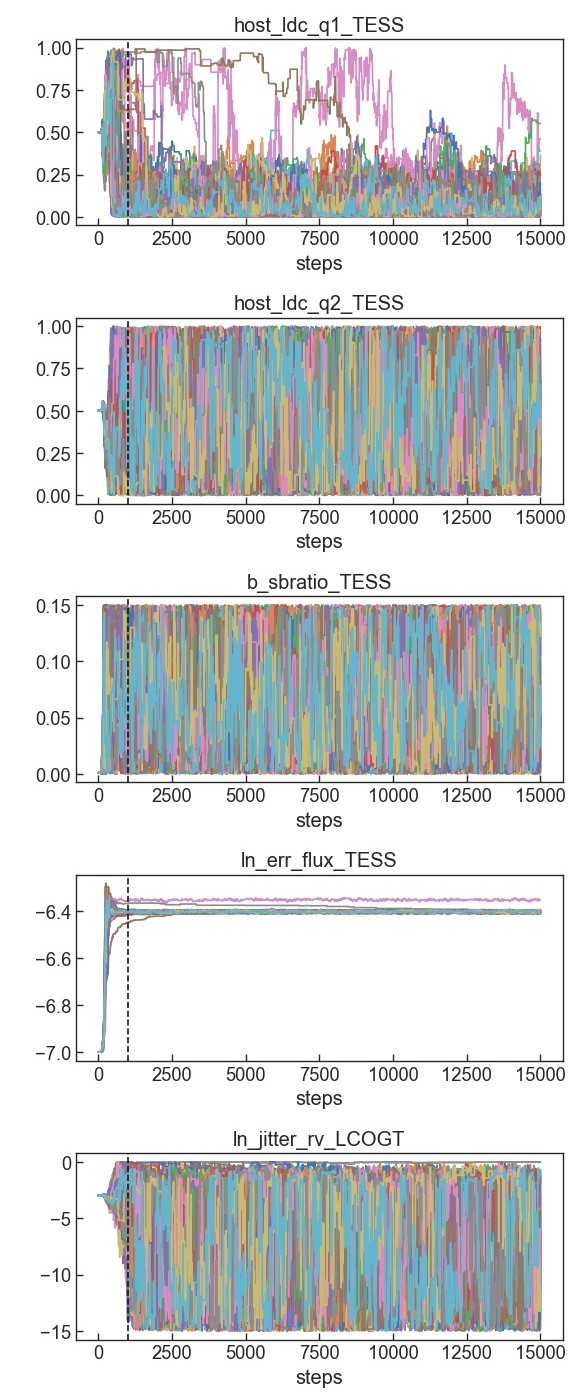}
     \end{subfigure}
    \caption{Chains for the \texttt{allesfitter} run. Showing, top to bottom: (\textit{Left column}) log probability function, ratio $R_2/R_1$, ratio $(R_1+R_2)/a$, inclination, epoch $T_0$; (\textit{Central column}) orbital period, RV semi-amplitude $K_1$, eccentricity parameters $f_c$ and $f_s$, {\it TESS} dilution factor; (\textit{Right column}) host limb darkening coefficients $q_1$ and $q_2$, {\it TESS} surface brightness ratio, and log errors in {\it TESS} and LCOGT data. The chains show the system has converged to a neat steady-state.}
    \label{app:mcmc:chains}
\end{figure*}

\clearpage

\section{Narrowed range spectral fitting results}\label{app:spectralfit}

\begin{figure*}
    \includegraphics[width=\textwidth]{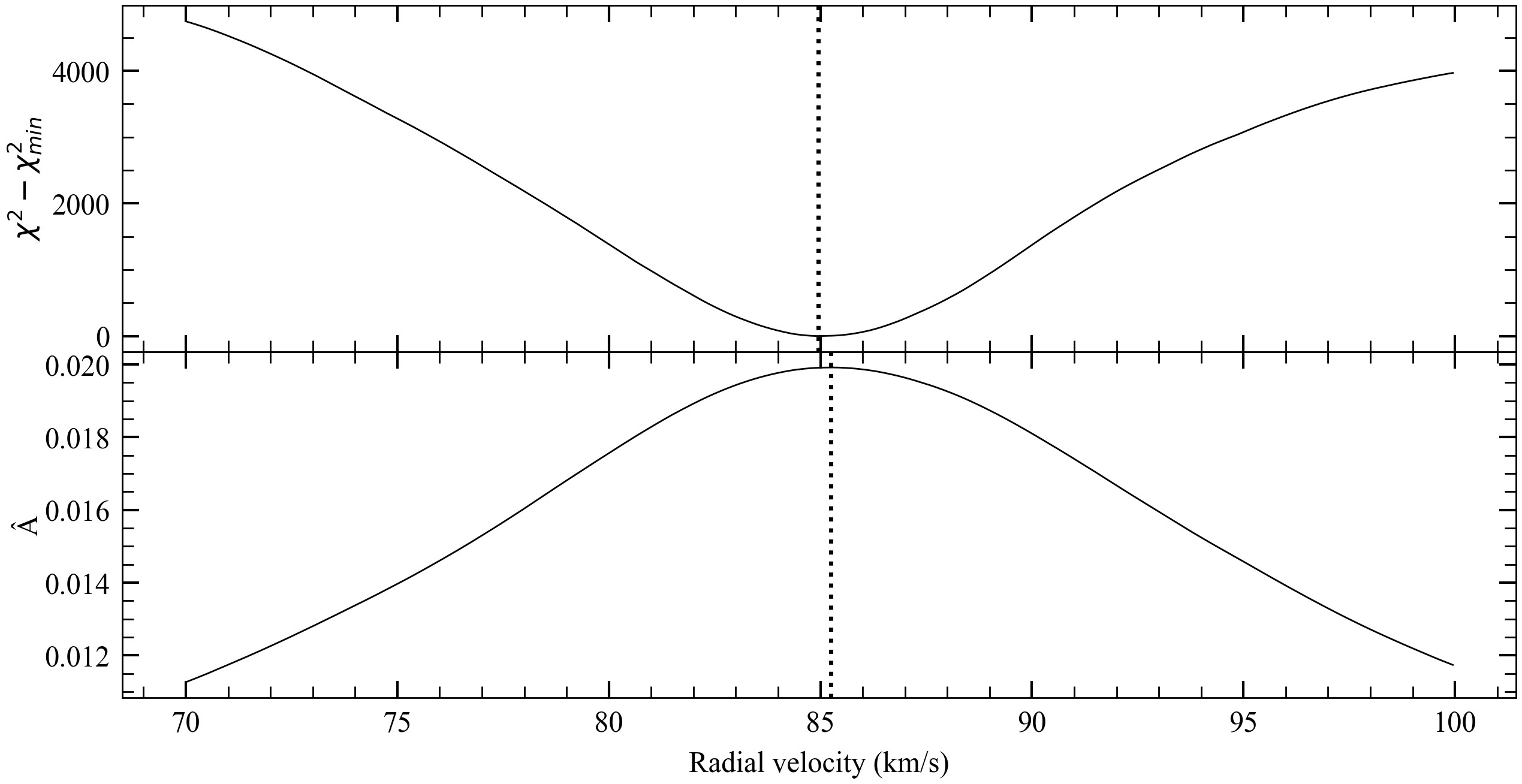}
    \caption{Results for chi-squared and optimal scaling factor with narrowed velocity grid range and increased grid resolution. Both peaks have remained in the vicinity of 85.5 km s\textsuperscript{-1} and it was thus suitable to use $\Delta \chi^2 = 1$ to determine the error bars.}
    \label{app:specfit:results}
\end{figure*}

\begin{figure}
    \includegraphics[width=\columnwidth]{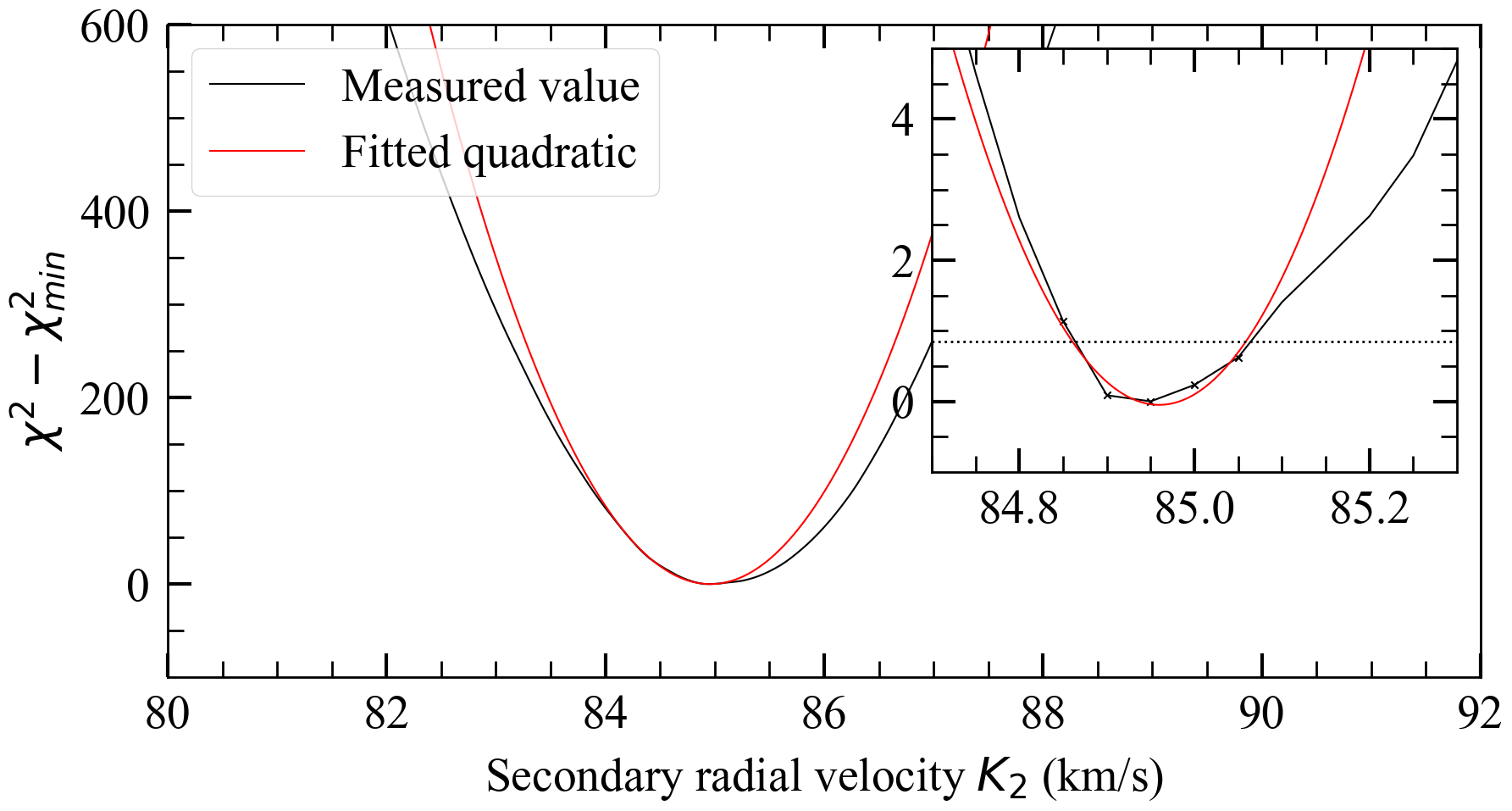}
    \caption{Fit for chi-squared minimum to determine error bars on increased velocity grid resolution.}
    \label{app:specfit:chimin}
\end{figure}

%%%%%%%%%%%%%%%%%%%%%%%%%%%%%%%%%%%%%%%%%%%%%%%%%%

% Don't change these lines
\bsp	% typesetting comment
\label{lastpage}
\end{document}